\documentclass[%
 reprint,
 amsmath,amssymb,
 aps,prd
]{revtex4-2}

\usepackage{graphicx}
\usepackage{dcolumn}
\usepackage{bm}
\usepackage{xcolor}

\RequirePackage{xspace} 
\newcommand{\sun}{$\mathop{\rm SU}(N_c)$\xspace}

\begin{document}

\title{Gauge covariant neural network for quarks and gluons}

\author{Yuki Nagai}%
 \email{nagai.yuki@mail.u-tokyo.ac.jp}
\affiliation{
Information  Technology  Center, 
The  University  of  Tokyo, 
6–2–3  Kashiwanoha,  Kashiwa,  Chiba  277–0882,  Japan
} 
\affiliation{%
CCSE, Japan Atomic Energy Agency, 178-4-4, Wakashiba, Kashiwa, Chiba 277-0871, Japan
}
\affiliation{
Mathematical Science Team, RIKEN Center for Advanced Intelligence Project (AIP), 1-4-1 Nihonbashi, Chuo-ku, Tokyo 103-0027, Japan
}

\author{Akio Tomiya}
\email{akio@yukawa.kyoto-u.ac.jp}
\affiliation{
Department of Information and Mathematical Sciences, Tokyo Woman’s Christian University, Tokyo 167-8585, Japan
}
\affiliation{
RIKEN/BNL Research center, Brookhaven National Laboratory, Upton, 11973, NY, USA
}
\affiliation{Faculty of Technology and Science, International Professional University of Technology, 3-3-1, Umeda, Kita-ku, Osaka, 530-0001, Osaka, Japan
}

\date{\today}

\begin{abstract}
%
We propose gauge-covariant neural networks along with a specialized training algorithm for lattice QCD, designed to handle realistic quarks and gluons in four-dimensional space-time. 
We show that the smearing procedure can be interpreted as an extended version of residual neural networks with fixed parameters.  
To demonstrate the applicability of our neural networks, we develop a self-learning hybrid Monte Carlo algorithm in the context of two-color QCD, yielding outcomes consistent with those from the conventional Hybrid Monte Carlo approach.
\end{abstract}

\maketitle

\section{Introduction}
\begin{figure}[h]
(a)
\begin{center}
\includegraphics[scale=0.12]{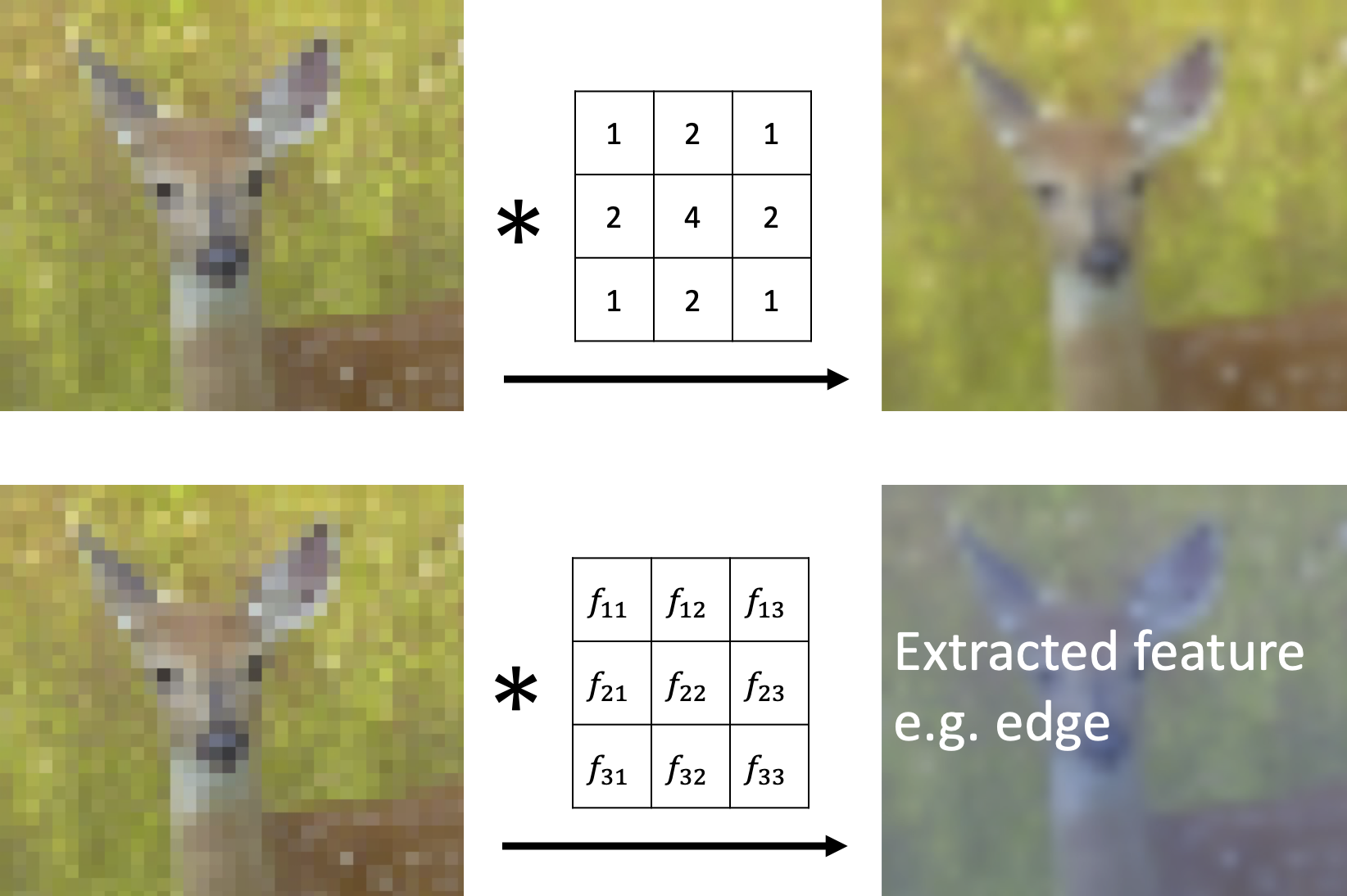}
\end{center}
(b)
\begin{center}
\includegraphics[scale=0.3]{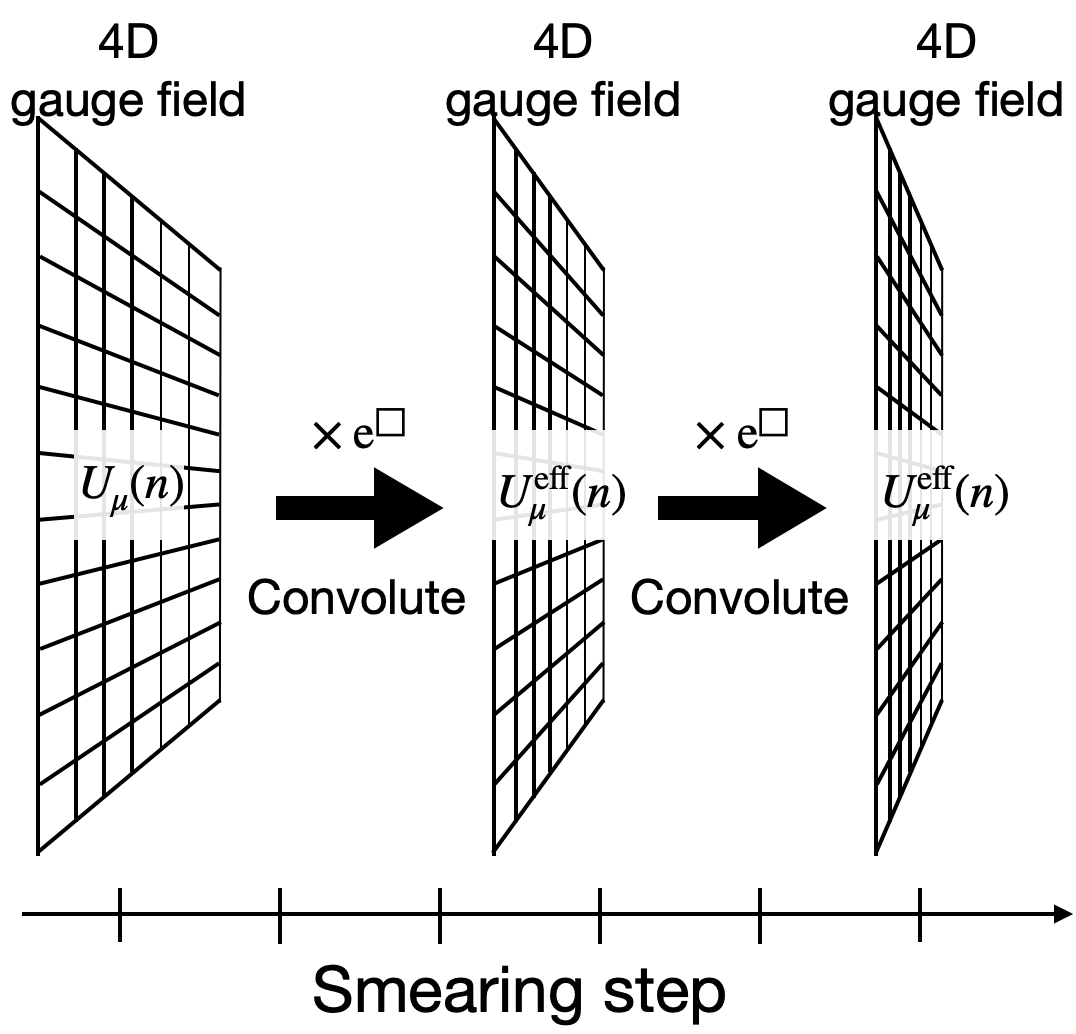}
\end{center}
\caption{
Schematic figures of filtering, CNN, and smearing. 
({\it Top} of a) Images are smoothened by the Gaussian filter ($3\times 3$).
({\it Bottom} of a) Images are fed in the convolutional layer, which is a
trainable filter, and the operation is determined by training.
(b) Schematic figure of the stout smearing \cite{Morningstar_2004}.
The gauge configuration is smoothened in a covariant manner.
\label{fig:pictrical_images}}
\end{figure}

Convolutional neural networks (CNNs) have become some of the most successful methods in machine learning, particularly in image recognition (Fig.\;\ref{fig:pictrical_images} (a) top) \cite{Akinori2021DLAP,bronstein2021geometric}.  
One reason for this success is the network’s structure, which ensures that the output remains invariant to shifts in the input data due to translational operations.  
When considering CNNs as filters on the input data, they can be regarded as trainable filters that respect this symmetry
 (Fig.\;\ref{fig:pictrical_images} (a) bottom).

Continuous symmetries are fundamental concepts in theoretical physics.  
The energy-momentum tensor, a crucial observable, is defined by Noether's theorem, which applies to global translational symmetry.  
Quarks and gluons, the fundamental building blocks of our universe, are described by a non-Abelian gauge-symmetric quantum field theory known as quantum chromodynamics (QCD).  
Gauge symmetry is essential for ensuring renormalizability.  
Lattice QCD is a highly successful formulation for calculating QCD, preserving gauge symmetry under a cutoff, reproducing the hadron spectrum, and predicting phase transition temperatures and other observables.
\cite{Ding:2015ona,Miura:2019xtd,Aoki:2019cca}. 

The smearing procedure \cite{Albanese:1987ds, Kamleh:2004xk, Morningstar_2004, Hasenfratz_2007}, a smoothing technique applied to gauge fields (Fig.\;\ref{fig:pictrical_images} (b)), has been widely used to reduce cutoff effects resulting from the finite lattice spacing in lattice QCD \cite{Blum:1996uf, Orginos:1999cr, Follana_2007, Bazavov_2010, Luscher:2013cpa}.  
We note that smearing can be interpreted as a filter that preserves the global translational, rotational, and \sun gauge symmetries in lattice QCD, where $N_c = 2, 3, 4, \dots$ denotes the number of colors.

We introduce a {\it trainable} filter that respects \sun gauge symmetry and global translational and rotational symmetries on the lattice, viewing smearing as a filter that maps between rank-2 tensor-valued vector fields.  
This trainable filter is applicable to systems with quarks and gluons in any dimension, given that smearing has been rigorously tested and studied over the past decades.  
While the input data in conventional CNNs consist of real-valued scalar data, the input data in the smearing filter consist of complex-valued \sun matrices.  
We refer to this filter as the {\it gauge covariant neural network}.  
As we will show, the covariant neural network can be trained using the extended delta rule, a scalar version of which is well known in the machine learning community.  
We find that the derivation of the extended delta rule for rank-2 tensor-valued vector fields parallels the well-known derivation of the Hybrid Monte Carlo (HMC) force for smeared fermions \cite{Morningstar_2004, Kamleh:2004xk}.  
In other words, with the extended delta rule, one can directly obtain derivatives of any quantities with respect to the bare gauge links, similar to the smeared force.

Several methods using machine-learning techniques have been developed for various applications in lattice QCD \cite{Tanaka:2017niz, Zhou:2018ill, Urban:2018tqv, Albergo:2019eim, rezende2020normalizing, Kanwar:2020xzo, Boyda:2020hsi, Albergo:2021vyo, Favoni:2020reg}.  
The most distinctive feature of our neural network is that it naturally incorporates dynamical fermions.  
In this paper, to demonstrate the potential applications of our neural network, we perform a simulation using self-learning hybrid Monte Carlo (SLHMC) \cite{Nagai_2020} on a non-Abelian gauge theory with dynamical fermions.  
SLHMC employs a parameterized fermion action in the molecular dynamics, and we use a gauge covariant neural network to parameterize the fermion action.

This paper is organized as follows:  
We introduce a neural network with covariant layers that can process non-Abelian gauge fields on the lattice while preserving symmetries, and we define gauge-invariant loss functions based on lattice actions.  
We then derive the delta rule for training the neural network, analogous to the derivation of the HMC force for smeared fermions.  
Next, we discuss the relationship between stout smearing and ResNet \cite{he2015deep, he2016identity}, one of the most prominent architectures in neural networks.  
Furthermore, we find that the neural ODE (ordinary differential equation) formulation of our network has the same functional form as the gradient flow \cite{L_scher_2009}.  
We develop the SLHMC with covariant neural networks to demonstrate potential applications, yielding results consistent with those from HMC.  
Finally, we summarize our findings in the concluding section.

\section{Effective model}
\subsection{Machine learning techniques in lattice QCD}
We will now discuss the importance of machine learning techniques in lattice QCD.  
Machine learning techniques are commonly employed to approximate functions $f \sim g$, where $f$ and $g$ represent the target function and the neural network approximation, respectively.

For example, in supervised learning, a neural network $g(x)$ is trained to reproduce an output variable $y = f(x)$ given an input variable $x$.  
In unsupervised learning, a neural network $g$ is used to approximate a probability distribution $f(x)$.

The self-learning Monte Carlo (SLMC) method, an application of supervised learning, is one of the most powerful approaches to accelerate Markov chain Monte Carlo (MCMC) simulations.  
In MCMC simulations, designing an efficient update method for proposing new configurations is crucial.  
In the lattice QCD community, the hybrid Monte Carlo (HMC) method has been widely used, where configurations are updated through molecular dynamics.  
SLMC employs an effective model to propose new configurations.

The authors have developed SLMC specifically for lattice systems with quarks and gluons.  
In this approach, configurations are updated using the heatbath method, employing a gluon-only effective action that approximates the full action with quarks and gluons.

The rapid development of generative models, particularly normalizing flows, has driven applications in lattice QCD.  
A normalizing flow is a type of neural network that represents mappings between probability distributions.  
The flow-based sampling algorithm utilizes normalizing flows to approximate mappings between a trivial distribution and the probability distribution of an interacting field theory.  
We can efficiently sample configurations from the trivial distribution and deform them into interacting configurations with straightforward parallelization.  
Furthermore, an exact distribution can be achieved by applying the Metropolis-Hastings test.  
This approach has been applied to both scalar field theory and QCD.

In both applications, the development of effective neural networks for lattice QCD is crucial.

\subsection{Required properties of neural networks for lattice QCD} \label{sec:gaugeqcd}
It is essential to carefully design neural network architectures for lattice QCD, despite the many types of neural networks available in the machine learning community.  
For example, in the SLMC application, a neural network can approximate the action:
\begin{align}
    S(\{U\}) = g(\{ U \}),
\end{align}
where $\{ U \}$ is a given configuration of the gauge fields in lattice QCD theory.  
To accelerate Monte Carlo simulations for lattice QCD, the proposed neural network should have the following properties: capability to handle both quarks and gluons, suitability for massively parallel computation, and preservation of gauge symmetry on the lattice.

Firstly, the proposed neural network should be able to treat both quarks and gluons.  
The computational complexity primarily arises from the treatment of quarks in lattice QCD simulations.  
Overall simulation speedup is achieved by reducing the computational cost of treating quarks.  

Secondly, massively parallel computation on supercomputers must be considered.  
Algorithms in lattice QCD must be implemented on supercomputers due to the high computational complexity involved in handling both quarks and gluons.  

Finally, the proposed neural network should preserve gauge symmetry on the lattice.  
Since the target action should be gauge invariant, the output of the neural network should also be gauge invariant:
\begin{align}
    g(\{ U\} ) = g(T[\{ U\}] ),
\end{align}
where $T$ is a gauge transformation.  
There are two ways to achieve symmetry in a neural network: constructing symmetry-invariant inputs or using covariant networks.  
With conventional neural networks, such as fully-connected neural networks that do not account for symmetry, constructing symmetry-invariant inputs is a feasible approach, expressed as
\begin{align}
    g(\{ U\} ) &= f_{\rm NN}(\vec{c}),\\ 
    \vec{c} &= g_{\rm input}(\{ U \}),
\end{align}
where $g_{\rm input}$ is a symmetric operation under a transformation $T$ ($g_{\rm input}(\{ U \}) = g_{\rm input}(T[\{ U\}])$).  
In lattice QCD, the traced plaquettes $\{ P(\{ U \}) \}$ for a given $\{ U \}$ can serve as the operation $g_{\rm input}$.  
Trainable parameters are included only in the function $f$, not in $g_{\rm input}$, as vector $\vec{c}$ serves as the input for the trainable neural network $f_{\rm NN}$.  

The other approach proposed in this paper involves constructing symmetry-covariant neural networks, formulated as
\begin{align}
    g(\{ U\} ) &= f(\{ U' \}),\\ 
    \{ U' \} &= g_{\rm NN}(\{ U \}).
\end{align}
Here, $f(\{ U'\})$ is a symmetry-invariant operation, such as calculating an action, satisfying $f(\{ U' \}) = f(\{ T[U'] \})$.  
If $g_{\rm NN}(\{ U \})$ is symmetry covariant, given by
\begin{align}
    g_{\rm NN}(\{ T[U] \}) = T \left[ g_{\rm NN}(\{ U \}) \right], \label{eq:gnn}
\end{align}
then the function $g(\{ U\} )$ is symmetry invariant ($g(\{ T[U]\} ) = g(\{ U\})$).  
The neural network $g_{\rm NN}(\{ U \})$ retains information about the symmetry, whereas this information is discarded in the symmetry-invariant inputs $\vec{c}$.

In this paper, we propose a new type of gauge-covariant neural network, which naturally extends the well-known theoretical method of smearing in lattice QCD.  
Smearing was developed to reduce unwanted fluctuations in the ultraviolet region of gauge fields, which can be viewed as a gauge-symmetric version of Gaussian smoothing operations used for images.  
It smooths nearby link variables and stabilizes the behavior of operators that include them.  
Smearing is equivariant under rotation and translation, and covariant under gauge transformations.

\subsection{Self-learning hybrid Monte Carlo method}
One of the applications of machine learning is the self-learning Hybrid Monte Carlo (SLHMC) method.  
The SLHMC is an exact algorithm developed in the condensed matter community \cite{Nagai_2020}, consisting of two components: the molecular dynamics part and the Metropolis test.  
We apply the SLHMC to lattice QCD by replacing the target action $S[U]$ in the molecular dynamics part of the HMC with the effective action $S_\theta[U]$, represented by our gauge-covariant neural networks.  
The exactness of the results is ensured by using $S[U]$ in the Metropolis test.

In lattice QCD calculations, gauge fields $\{ U \}$ and the pseudofermion field $\{ \phi \}$ are distributed according to the Boltzmann weight factor:
\begin{align}
    \exp (-S[\phi,U]) = \exp (-S_g([U]) - S_f([\phi,U] )),
\end{align}
where $S_g[U]$ is the gauge action and $S_f[\phi,U]$ is the fermion action involving the pseudofermion field.  
In conventional HMC, molecular dynamics with forces derived from $S[\phi,U]$ is employed to propose new configurations $\{ U \}$.  
In SLHMC, we use forces derived from the effective model $S_{\rm eff}[\phi,U]$.  
In the Metropolis sampling test, the acceptance ratio is calculated based solely on the original action $S[\phi,U]$.  
Since the Metropolis test is based on the original action $S[\phi,U]$, the results are as accurate as those obtained by conventional HMC.  
If the effective model $S_{\rm eff}[\phi,U]$ provides a good approximation to the original action $S[\phi,U]$, the acceptance ratio during simulations is high.  
The computational complexity depends on the cost of evaluating the effective action $S_{\rm eff}[\phi,U]$.  
Therefore, if the computational complexity of $S_{\rm eff}[\phi,U]$ is low, the total computational complexity of the SLHMC is reduced.

In lattice QCD, the computational cost of dynamical fermions strongly depends on the types of fermions and quark mass.  
Several studies have shown that expectation values of observables for numerically expensive fermions can be calculated through reweighting techniques applied to less numerically expensive fermions \cite{Fukaya:2013vka, Tomiya:2016jwr}.  
In this paper, by applying SLHMC, we can employ relatively heavier fermions to generate configurations for systems with lighter fermions.

\section{Gauge covariant neural networks}

\subsection{Overview and structure}
Here, we introduce effective gauge fields constructed using a gauge-covariant neural network for lattice gauge theory.  
We define a link variable $U_{\mu}(n)$ representing gauge fields at a point $n$ on a four-dimensional lattice, with direction $\mu=1,2,3,4$.  
As discussed in Sec.~\ref{sec:gaugeqcd}, the effective gauge field $U^{\rm eff}_{\mu}(n)$ should be gauge covariant:  
\begin{align}
U^{\rm eff}_{\mu}(n) = g_{\rm NN}(\{ U \} )_{\mu}(n).
\end{align}
Here, we call $g_{\rm NN}(\{ U \} )$ a gauge covariant map. 
Since the gauge covariant map can be applied to effective gauge fields recursively, we consider the following effective gauge fields:
\begin{align}
    U^{(1)}_{\mu}(n) &= g_{\rm NN}(\{ U \} )_{\mu}(n), \\
    U^{(2)}_{\mu}(n) &= g_{\rm NN}(\{ U^{(1)} \} )_{\mu}(n),\\
    & \vdots \nonumber \\
    U^{(L)}_{\mu}(n) &=  g_{\rm NN}(\{ U^{(L-1)} \} )_{\mu}(n) \equiv U^{\rm eff}_{\mu}(n).
\end{align}

\begin{figure}[th]
\centering
\begin{tabular}{cc}
\includegraphics[scale=0.35]{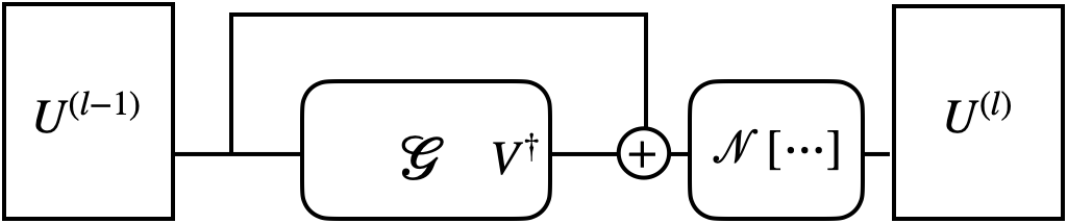}
\end{tabular}
\caption{
Calculation graph notation of neural networks for generalized smearing procedure.
Input and output of this component are matrices.
\label{fig:cov-net}}
\end{figure}

The gauge-covariant map $g_{\rm NN}$ can be considered a natural extension of the well-known theoretical method of smearing.  
As shown in Fig.~\ref{fig:cov-net}, the general form of the $l$-th smearing-type gauge-covariant map is defined as  
\begin{align}
z_\mu^{(l)}(n)&=
w_1^{(l-1)} U^{(l-1)}_{\mu}(n) 
+ w_2^{(l-1)} {\cal G}_{\mu,n}^{\bar{\theta}^{(l-1)}}[ U^{(l-1)} ]
\label{eq:generalized_smear1},\\
U^{(l)}_{\mu}(n) &= {\cal N} (z_\mu^{(l)}(n)) \label{eq:generalized_smear2}.
\end{align}
Here, ${\cal G}^{\bar{\theta}^{(l-1)}}_{\mu,n}[ U^{(l-1)} ]$ is a filter function that obeys the same gauge transformation law as a link $U^{(l-1)}_{\mu}(n)$.  
${\bar{\theta}^{(l-1)}}$ is a set of parameters in it.  
${\cal N}$ is an activation function that acts locally.  
For example, in the case of APE (HYP)-type smearing  
(See appendix \ref{sec:connect}) \cite{Albanese:1987ds,Hasenfratz_2007}, ${\cal G}^{\bar{\theta}^{(l-1)}}_{\mu,n}[ U^{(l-1)} ]$ is a linear function that consists of a sum of staples and ${\cal N}$ is a non-linear function to project (normalize) to \sun ( U($N_c$) ).  
In the case of stout-type smearing  
\cite{Morningstar_2004,Capitani:2006ni}, ${\cal G}^{\bar{\theta}^{(l-1)}}_{\mu,n}[ U^{(l-1)} ]$ is an exponential function of traceless-antihermitian closed loops, and ${\cal N}$ is the identity.  
In general, we can use any filter consisting of gauge-covariant combinations of links.  
$w_1^{(l-1)}, w_2^{(l-1)} $ are real parameters (weights in neural networks).  
They can be promoted to vector fields,  
$w_i^{(l-1)}\to w_{i,\mu}^{(l-1)}(n)$,  
and the current formulation corresponds to the weight-shared version.  
To guarantee rotational and translational symmetries on the lattice, we adopt weight sharing.  
As in conventional smearing, layers can be applied repeatedly.  
In other words, multi-step smearing \cite{Morningstar_2004} corresponds to neural networks with multiple layers ({\it i.e.}, deep neural networks) with fixed parameters.

\begin{figure}[t]
\begin{center}
\includegraphics[scale=0.5]{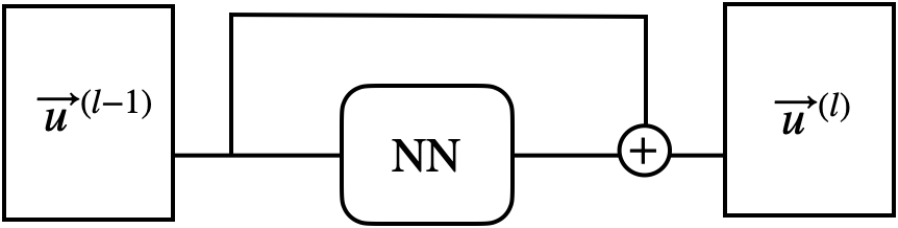}\vspace{3mm}
\end{center}
\caption{
Calculation graph notation of Res-Net.
Input and output of this component are vectors.
NN in the figure indicates a neural network.
\label{fig:cov-net_stout}}
\end{figure}

\subsection{Rank-2 residual-net (Res-Net)}
We show that the $l$-th gauge covariant map $U^{(l)}_{\mu}(n) = g_{\rm NN}(\{ U^{(l-1)} \} )_{\mu}(n)$ can be seen as a rank-2 tensor version of the residual network (Res-Net \cite{he2015deep, he2016identity}), as shown in Fig. \ref{fig:cov-net} and Fig. \ref{fig:cov-net_stout}, by setting $w_1^{(l-1)} = 1$ and ${\cal N}(x) = x$:  
\begin{align}
    U_\mu^{(l)}(n)&=  w_2^{(l-1)} {\cal G}_{\mu,n}^{\bar{\theta}^{(l-1)}}( U^{(l-1)} ) +  U^{(l-1)}_{\mu}(n) \label{eq:resnet}.
\end{align}
If $U$ consists of scalar values, this network structure is known as the Res-Net.  
Thus, we refer to Eq.~\eqref{eq:resnet} as the rank-2 Res-Net, whose components form a rank-2 tensor field.  
Therefore, our gauge covariant neural network can be regarded as a rank-2 tensor version of the multi-layer perceptron (See, appendix \ref{sec:full}). 

\subsection{Neural ordinary differential equation (Neural ODE)}
We can also introduce the rank-2 tensor version of the neural ODE \cite{chen2019neural}, a generalization of the conventional Res-Net, which is a cutting-edge framework in deep learning \cite{he2015deep, he2016identity}.  
If the layer index $l$ is regarded as a fictitious time $t$,  
Eq. \eqref{eq:resnet} leads to the rank-2 neural ODE,  
\begin{align}
\frac{d U_\mu^{(t)}(n)}{dt}
  &=  w_2^{(t)} {\cal G}_{\mu,n}^{\bar{\theta}^{(t)}}( U^{(t)} ). \label{eq:gradient_flow}
\end{align}
Conversely, one can derive the rank-2 Res-Net (\ref{eq:resnet}) by solving the differential equation above using the Euler method ($y_{n+1} = y_n + hf(t_n,y_n)$, for small $h$).  
Since a continuous version of stout smearing \cite{Luscher:2010iy} is known as the Wilson flow, a gradient flow with Wilson plaquette action, the flow with given parameters in lattice QCD can be regarded as a neural ODE for a gauge field within the framework of machine learning.

\subsection{Loss function and back propagation}
Neural networks are trained to minimize a scalar quantity known as a loss function in supervised learning.  
The loss function in lattice QCD, a functional of the effective gauge fields ${\cal L}(\{ U^{\rm eff} \})$, becomes minimal when its derivative with respect to trainable parameters is zero:  
\begin{align}
    \frac{\partial {\cal L}(\{ U^{\rm eff} \})}{\partial \theta} = 0.
\end{align}
For example, in SLHMC, we consider the mean-squared loss function defined as  
\begin{align}
    {\cal L}(\{ U^{\rm eff} \}) = \sum_{i} |S(\{ U_i \}) - S^{\rm eff}(\{ U^{\rm eff}_i \})|^2, 
\end{align}
where $\{ U_i \}$ is the $i$-th configuration.  
To minimize the loss function, we require the derivative of the scalar function (e.g., the action) with respect to trainable parameters.  
In addition, the derivative of the scalar function with respect to the gauge fields, defined as  
\begin{align}
\frac{\partial f(\{ U^{\rm eff}(\{U \}) \})}{\partial [U_{\mu}(n)]_{\alpha \beta}} \equiv \left[ \frac{\partial f(\{ U^{\rm eff}(\{U \}) \})}{\partial U_{\mu}(n)} \right]_{\beta \alpha },
\end{align}
is important for HMC simulations.  
Since $f$ is a functional of the effective gauge fields $U^{\rm eff}$, the above derivative with respect to the bare gauge fields $U$ cannot be directly obtained.  
In the machine learning community, automatic differentiation using the backpropagation technique is well known (see Appendix \ref{sec:fcnn}).

In this paper, we propose a rank-2 tensor version of backpropagation.  
The derivative of the scalar function $f$ with respect to the $l$-th effective gauge field $U_{\mu}^{(l)}(n)$ is expressed as  
\begin{align}
    \frac{\partial f}{\partial [U_{\mu}^{(l)}(n)]_{\alpha \beta}} &= \sum_{\mu,n,\alpha' \beta'}  \left[ 
     \frac{\partial f}{\partial  [z^{(l)}_{\mu'}(n')]_{\alpha' \beta'}} \frac{\partial  [z^{(l)}_{\mu'}(n')]_{\alpha' \beta'}}{\partial [U_{\mu}^{(l)}(n)]_{\alpha \beta}} \right. \nonumber \\
     & \left. + 
     \frac{\partial f}{\partial  [z^{(l)\ast}_{\mu'}(n')]_{\alpha' \beta'}} \frac{\partial  [z^{(l)\ast}_{\mu'}(n')]_{\alpha' \beta'}}{\partial [U_{\mu}^{(l)}(n)]_{\alpha \beta}}  \right], \\
    &= \sum_{\mu,n,\alpha' \beta'}
    \left[   [ \frac{\partial f}{\partial z^{(l)}_{\mu'}(n')} ]^{\beta'}{}_{\alpha'}
    [ \frac{\partial  z^{(l)}_{\mu'}(n')}{\partial U_{\mu}^{(l)}(n)} ]^{\alpha'}{}_{\alpha} {}^{\beta}{}_{\beta'}
    \right. \nonumber \\
&    \left. +
    [ \frac{\partial f}{\partial z_{\mu'}^{(l)\dagger}(n')} ]^{\alpha'}{}_{\beta'}
    [ \frac{\partial  z_{\mu'}^{(l)\dagger}(n')}{\partial U_{\mu}^{(l)}(n)} ]^{\beta'}{}_{\alpha} {}^{\beta}{}_{\alpha'}
    \right],
\end{align}
\begin{align}
\frac{\partial f}{\partial U_{\mu}^{(l)}(n)} 
    &= \sum_{\mu,n} \left[
    \frac{\partial f}{\partial z^{(l)}_{\mu'}(n')} \star  \frac{\partial  z^{(l)}_{\mu'}(n')}{\partial U_{\mu}^{(l)}(n)} \nonumber \right. \\
& \left. +
     \frac{\partial f}{\partial z_{\mu'}^{(l)\dagger}(n')} 
    \star  \frac{\partial  z_{\mu'}^{(l)\dagger}(n')}{\partial U_{\mu}^{(l)}(n)} \right],
\end{align}
Here, we use definition of matrix derivative on a scalar function $f$ as 
\begin{align}
     [\frac{\partial f}{\partial A}]^i{}_j \equiv \frac{\partial f}{\partial A^j{}_i}
\end{align}
, the rank-4 tensor $\partial M/\partial A$ is defined as 
\begin{align}
    [\frac{\partial M}{\partial A}]^i{}_j {}^k{}_l &\equiv \frac{\partial }{\partial B^j{}_k} [M]^i{}_l
\end{align}
and the star product for the rank-2 and rank-4 tensors defined as 
\begin{align}
    [A \star T]^i{}_j &\equiv \sum_{kl}  A^l{}_k T^k{}_j{}^i{}_l,
\end{align}
which is originally introduced in Ref.~\cite{Kamleh:2004xk} (in detail, see appendix \ref{sec:tensor}). 
We introduce the rank-2 {\it local} derivative $\delta_{\mu}^{(l)}(n)$ defined as 
\begin{align}
   \delta_{\mu}^{(l)}(n) \equiv  \frac{\partial f}{\partial z^{(l)}_{\mu'}(n')},
\end{align}
$\delta^{(L)}_{\mu}(n)$ on the final layer $l = L$ becomes 
\begin{align}
    \delta^{(L)}_{\mu}(n) &= \sum_{\alpha=I,II}  \frac{\partial S_\theta}{\partial  U^{(L)\alpha}_{\mu}(n)} \star
     \frac{\partial {\cal N} [  z^{(L)}_{\mu}(n)]^{\alpha}}{\partial  z^{(L)}_{\mu}(n)}  ,
\end{align}
Here, we define $A^{I}(n) \equiv A$ and  $A^{II} \equiv A^{\dagger}$. 
On the $l$-th layer,  $\delta^{(l)\alpha}_{\mu}(n)$ is written as 
\begin{align}
  &   \delta^{(l)}_{\mu,\alpha}(n) 
    =  \sum_{\mu',m,\beta} 
     \delta^{(l+1)}_{\mu',\beta}(m)
     \star \frac{\partial z^{(l+1)\beta}_{\mu'}(m)}{\partial  z^{(l)\alpha}_{\mu}(n)}, \\
 &  
     = \sum_{\beta=I,II} \Bigg\{ 
      w_1^{(l)} \delta^{(l+1)}_{\mu,\beta}(n)      \star 
     \frac{\partial  
           {\cal N} [  z^{(l)\beta}_{\mu}(n)]
          }{\partial z^{(l)\alpha}_{\mu}(n) }  \nonumber \\
&    +  w_2^{(l)}   \sum_{\mu',m} 
      \delta^{(l+1)}_{\mu',\beta}(m) \nonumber \\
      &\star  \sum_{\gamma=I,II} 
       \frac{\partial  {\cal G}_{\mu',m}( U^{(l)} )^{\beta}}{\partial U^{(l)\gamma}_{\mu}(n) } 
       \star  \frac{\partial  {\cal N} [  z^{(l)}_{\mu}(n)]^{\gamma}}{\partial z^{(l)\alpha}_{\mu}(n) }     
       \Bigg\},
\end{align}
where $\alpha,\gamma = I, II$.
This represents the rank-2 delta rule for the covariant neural network (for details, see Appendix \ref{sec:rank2delta}).  
Using this delta rule, we can optimize the network weights through a gradient optimizer.

\section{Demonstration}
Here, we demonstrate how the gauge covariant neural network can be applied.  
We conduct simulations of the self-learning hybrid Monte-Carlo (SLHMC) \cite{Nagai_2020} for non-Abelian gauge theory with dynamical fermions as a demonstration.  
In SLHMC, the action of the target system is used in the Metropolis test, while a machine learning-approximated action is used in the molecular dynamics.  
We apply a gauge-covariant neural network-approximated action in the SLHMC for gauge theory.

\subsection{Self-learning Hybrid Monte-Carlo for QCD}
SLHMC is an exact algorithm consisting of two parts: the molecular dynamics and the Metropolis test.  
We use $S[U]$, the action for the target system, in the Metropolis test, and $S_\theta[U]$, an effective action for the system with a set of parameters $\theta$, in the molecular dynamics.  
Since the molecular dynamics is reversible, performing the Metropolis-Hastings test is unnecessary. Instead, the Metropolis test alone is sufficient to guarantee convergence \cite{Nagai_2020, Nagai:2020jar}.  
The same mechanism has been used in the rational hybrid Monte-Carlo \cite{Clark:2006wq}.

Our target system is given by  
\begin{align}
S[U] = S_\text{g}[U] + S_\text{f}[\phi, U;m_\text{l}],
\end{align}
where $S_\text{g}[U]$ is a gauge action and 
$S_\text{f}[\phi, U;m_\text{l}]$ is the action for a pseudo-fermion field $\phi$ with mass $m_\text{l}$.  
In HMC, the total action is a function of the pseudo-fermion field, but here we suppress it in the equation for notational simplicity.

In SLHMC for QCD, we can take the following effective action in general,  
\begin{align}
S_\theta[U] = S_\text{g}^\text{eff}\big[U^\text{NN}_{\theta'}[U]\big] + S_\text{f}^\text{eff}\big[\phi, U^\text{NN}_\theta[U];m_\text{h}\big],
\end{align}
where $U^\text{NN}_\theta[U]$ is a transformed configuration from a trained covariant neural network, and $U^\text{NN}_{\theta'}[U]$ may represent another network.  
$m_\text{h}$ is the fermion mass in the effective action, which is chosen such that $m_\text{h}>m_\text{l}$.  
$S_\text{g}^\text{eff}$ is a gauge action, which may contain a tunable parameter as in \cite{Nagai:2020jar}, but here we simply use the Wilson plaquette action.  
$ S_\text{f}^\text{eff}$ is also a fermion action and may differ from the target action, and we will mention this extension, but we use the same action as the target, except for the gauge field in the fermion action. Specifically, we take  
\begin{align}
S_\theta[U] = S_\text{g}\big[U\big] + S_\text{f}\big[\phi, U^\text{NN}_\theta[U];m_\text{h}\big],
\end{align}
in our calculation. This is a neural network-parameterized effective action.  
If the neural network has sufficient expressivity, as shown in \cite{cybenko1989approximation, zhou2018universality, kidger2020universal, tabuada2020universal, johnson2018deep}, we expect that the effective action can approximate the target one as a functional of $U$ and $\phi$.


We employ stout-type smearing with parameters $w_1^{(l)} = w_2^{(l)} = 1$, ${\cal N}(z) = z$, and ${\cal G}_{\mu,n}( U^{(l)}) = \exp (Q_{\mu}^{(l)}(n)) - 1$ (see Appendix \ref{sec:stout}).  
Here, $Q_{\mu}^{(l)}(n)$ is defined as $2[\Omega_{\mu}^{(l)}(n)]_\text{TA}$, where $\text{TA}$ denotes the traceless-antihermitian operation.  
In conventional stout-type smearing, $\Omega^{(l)}_{\mu}(n)$ consists of untraced plaquette loop operators \cite{Morningstar_2004}.  
In SLHMC, one can consider any kind of closed loop operators, such as HEX smearing \cite{Capitani:2006ni}.  
In this paper, we define $\Omega^{(l)}_{\mu}(n)$ with trainable parameters as  
\begin{align}
\Omega^{(l)}_{\mu}(n) 
&= 
\rho^{(l)}_\text{plaq} O^\text{plaq}_{\mu}(n) 
\notag\\
&+
\begin{cases}
\rho^{(l)}_\text{poly,s} O^\text{poly}_{i}(n) ,
\;\;\text{($\mu=i=1,2,3$)},\\
\rho^{(l)}_\text{poly,$4$} O^\text{poly}_{4}(n), 
\;\;\text{($\mu=4$)}.
\end{cases}
\label{eq:omega_stout_train_practice}
\end{align}
where $O^\text{plaq}_{\mu}(n)$ and $O^\text{poly}_{\mu}(n)$ represent the plaquette and Polyakov loop operators in the $\mu$ direction, originating from a point $n$.  
For symmetry, we assign identical weights to spatial Polyakov loops, $\rho^{(l)}_\text{poly,s}$.  
The weights $
\rho^{(l)}_\text{plaq},
\rho^{(l)}_\text{poly,s},
\rho^{(l)}_\text{poly,$4$}
\in \mathbb{R}$ are determined during training.  
We set the number of network layers as $L=2$  
($
U^\text{NN}_\theta[U]
=
U^{(l=2)}[U^{(l=1)}[U]
]$).
The detailed derivation of the backpropagation for the stout smearing is shown in Appendix \ref{sec:stout}. 

In our previous work \cite{Nagai:2020jar}, we had to use the fermion determinant directly, as SLMC for QCD requires the absolute value of the fermionic determinant during training and production.  
However, in the current work, we do not have to use the determinant directly, as it relies on the SLHMC framework.  
Note that our choice in Eq.~\eqref{eq:omega_stout_train_practice} is a natural extension of the effective action from our previous work \cite{Nagai:2020jar}.

\subsection{Training}
Here we describe how to train our neural network.  
We train $S_\theta$ during an HMC run prior to the SLHMC run.  
After the Metropolis step in HMC, we retain the pseudo-fermion field and gauge configuration $U$ and calculate  
\begin{align}
L_\theta[U] = \frac{1}{2}\Big|S_\theta[U,\phi]-S[U,\phi] \Big|^2,
\end{align}
and perform backpropagation using the delta rule.  
This is a gauge-invariant loss function.  
For simplicity, we generate gauge configurations using the trained neural network after completing training. Training and production can also be performed simultaneously as in the previous paper \cite{Nagai:2020jar}.

After training, we generate configurations using the neural network.  
As explained above, we use the target action $S$ in the Metropolis test in SLHMC and $S_\theta$ in the molecular dynamics.  
For simplicity, we do not change the weights during the SLHMC run.

\subsection{Numerical setup}
Here, we summarize our numerical setup.  
Our framework can be implemented immediately if a public code includes smearing functionality.  
We implement it for SU(2) gauge theory in four dimensions with dynamical fermions.  
Our code is implemented based on LatticeQCD.jl \cite{LatticeQCDjl} in Julia \cite{bezanson2015julia}.  
We use an automatic staple generator from loop operators, which is available in our public code \cite{LatticeQCDjl}.  
In addition, we implement an automatic staple derivative generator for arbitrary general staples in the delta rule.

We perform simulations with unrooted staggered fermions and the plaquette gauge action in a $N_\sigma^3 N_\tau=4^4$ lattice with $m_\text{l}=0.3$ and $\beta = 2.7$ for this proof-of-principle study.  
We train our network with $m_\text{h}=0.4, 0.5, 0.75, 1.0$ (see Tab. \ref{tab:setup_train}) to observe the training history of the loss function and weights.

We choose a training rate of $\eta = 1.0\times 10^{-7}$, which is not precisely tuned but sufficient for this demonstration, and we employ simple stochastic gradient descent as the optimizer.  
Training was performed over 1500 trajectories (training steps) starting from a thermalized configuration.

The SLHMC is performed using trained weights and $m_\text{h}=0.4$.  
To increase statistical accuracy, the simulation is performed with multiple streams using different random seeds, and the first 1000 trajectories of each stream are excluded from the analysis.  
For SLHMC, we use 50000 trajectories in total.  
The step size in the molecular dynamics is set to 0.02 to evaluate the quality of the trained action.  
For HMC, we use the same number of trajectories for analysis.

\begin{table}[htb]
\centering
\begin{tabular}{ccccc} 
$N_\tau$ &$N_\sigma$ & $m_\text{h}$  & $N_\text{layers}$ & Loops \\
\hline \hline
4&4 & 1.0 & 2& Plaquette, Polyakov loops\\
4&4 & 0.75 & 2& Plaquette, Polyakov loops\\
4&4 & 0.5 & 2& Plaquette, Polyakov loops\\
4&4 & 0.4 & 2& Plaquette, Polyakov loops\\
\end{tabular} 
\caption{``Loops'' means types of untraced loop operator in the stout-type covariant neural networks.
In each case, the network contains 6 tunable weights.
Polyakov loops contain both of spatial and temporal loops.
\label{tab:setup_train}
}
\end{table}

\section{Results}
\subsection{Effective mass dependence}
\begin{figure}[t]
\begin{center}
\includegraphics[scale=0.55]{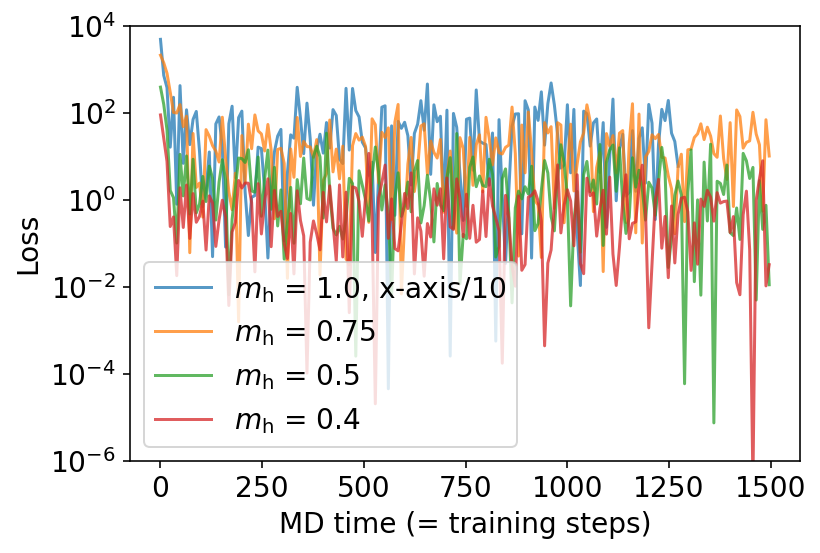}
\includegraphics[scale=0.55]{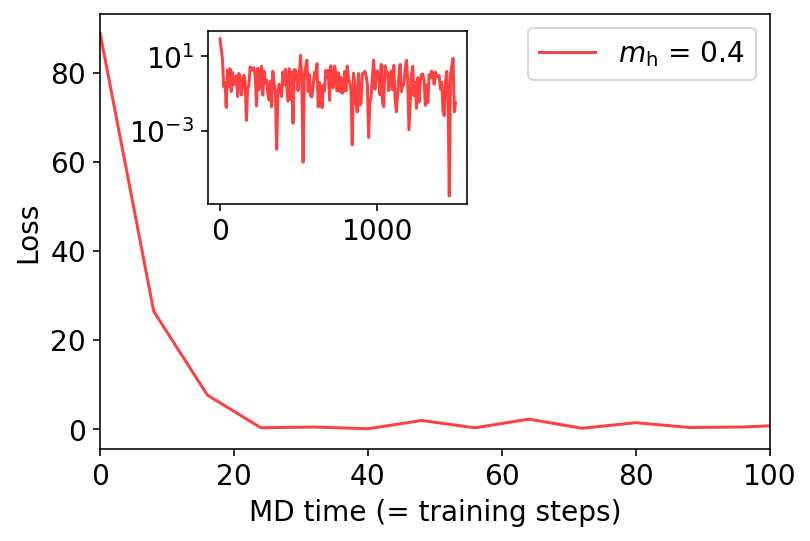}
\end{center}
\caption{
Loss history in prior run of HMC.
Horizontal axis coincides with the training steps.
({\it Top})
Summary plot for $m_\text{h}=1.0, 0.75, 0.5,0.4$ in the log scale.
({\it Bottom})
Loss history for $m_\text{h}=0.4$ in the linear scale. The inside plot is the same one except for the range of the horizontal axis.
\label{fig:loss_history}}
\end{figure}

We numerically demonstrate that the effective action produced by our gauge-covariant neural networks can approximate the target action.  
The top panel of Fig. \ref{fig:loss_history} shows the loss function history for various $m_\text{h}$ values on a log scale, along with molecular dynamics time from the preceding HMC, which coincides with the training steps in this case.  
For $m_\text{h}=1.0$, convergence is not achieved around a molecular dynamics (MD) time of 1500, so we extend the training to more than 10,000 steps.  
In cases where $m_\text{l} \sim m_\text{h}$, the loss functions fluctuate around 1.  
For $m_\text{h} \gg m_\text{l}$, it starts around $O(10^3)$ and decreases to $O(1)$.  
These results indicate that the neural network is correctly trained using stochastic gradient descent and our formulas.  
The bottom panel of Fig. \ref{fig:loss_history} shows the loss history for $m_\text{h}=0.4$ on a linear scale. The inset shows a zoomed-in view around MD time 0.  
In the early stage of training, it starts at 89 and decreases with fluctuations. After some steps, it fluctuates around $O(1)$.
The mass dependence of the trainable parameters is shown in Appendix \ref{sec:mass}. 

\subsection{SLHMC as an exact algorithm}
\begin{figure}[th]
\begin{center}
\includegraphics[scale=0.5]{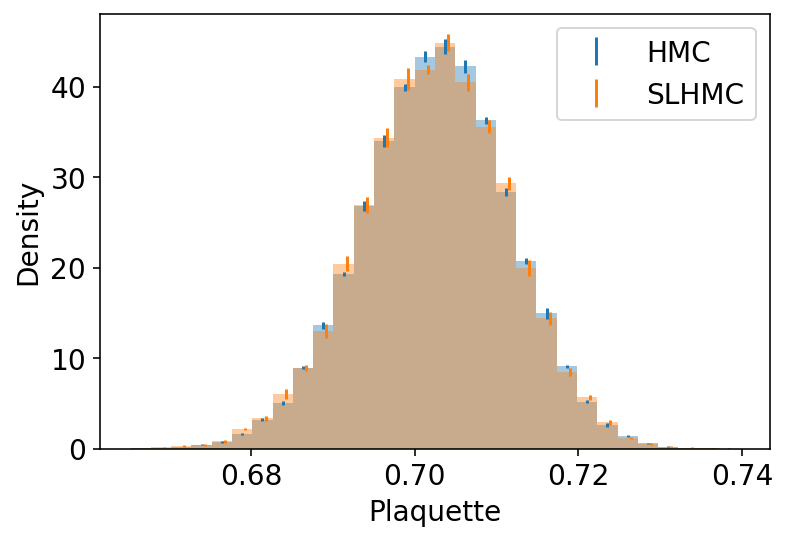}
\includegraphics[scale=0.5]{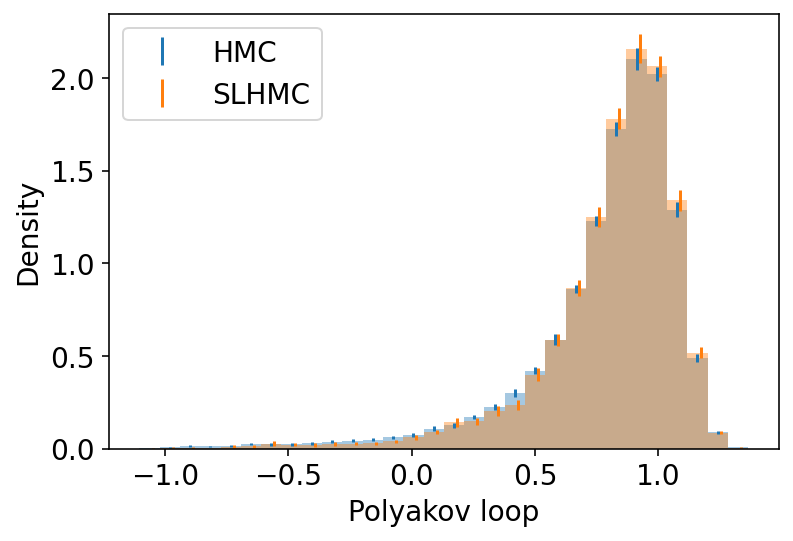}
\includegraphics[scale=0.5]{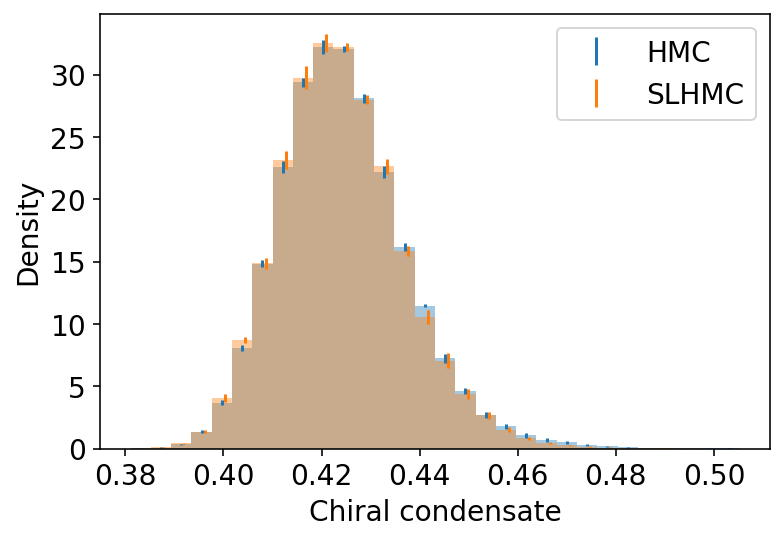}
\end{center}
\caption{
Comparison of results for several observables by HMC and SLHMC.
Error bars for SLHMC are horizontally shifted to right to avoid overlap of symbols.
(Top) plaquette, 
(Middle) Polyakov loop for the temporal direction.
(Bottom) the chiral condensate.
\label{fig:histogram}}
\end{figure}

Fig. \ref{fig:histogram} shows histograms of the observables, plaquette, and Polyakov loops, obtained from HMC and SLHMC.  
Error bars are calculated using the Jackknife method. 
All histograms show good overlap.  
Table \ref{tab:results} presents a summary of the results from HMC and SLHMC.  
Autocorrelation effects on the statistical error are taken into account.  
It can be observed that the two algorithms yield consistent results within the error margins.

\begin{table}[htb]
\centering
\begin{tabular}{cc|c} 
Algorithm &         Observable &       Value \\
\hline \hline
      HMC &          Plaquette &  0.70257(7) \\
    SLHMC &          Plaquette &   0.7024(1) \\
    \hline
      HMC &    $\|\text{Polyakov loop}\|$ &    0.812(7) \\
    SLHMC &    $\|\text{Polyakov loop}\|$ &    0.824(9) \\
    \hline
      HMC &  Chiral condensate &   0.4245(3) \\
    SLHMC &  Chiral condensate &   0.4240(3) \\
\end{tabular}
\caption{
Summary table for results from HMC and SLHMC.
\label{tab:results}
}
\end{table}

\subsection{Acceptance ratio}
Finally, we provide a comment on the acceptance rate in this demonstration.  
The initial value of the loss function is approximately 90, indicating that the acceptance rate without training would be nearly zero, as $ \mathrm{e}^{-L_\theta} \sim \mathrm{e}^{-90}$.  
It should be noted that only six parameters are used in this demonstration.  
The number of parameters can be increased by adding more layers, different types of loop operators, and other enhancements.

\section{Summary and discussion}
We introduce gauge-covariant neural networks that create mappings between gauge configurations while preserving gauge covariance.  
Our neural networks are constructed by setting the parameters of conventional smearing techniques as trainable.  
Using the network output, we define a real-valued, gauge-invariant loss function based on gauge-invariant actions.  
We note that the covariant network can be regarded as a rank-2 version of ResNet, with its neural ODE exhibiting a structure similar to L\"uscher's gradient flow.  
The covariant neural network can be trained using the rank-2 delta rule, which we also develop in this paper.

We apply SLHMC \cite{Nagai_2020} to two-color QCD with dynamical fermions in four dimensions to demonstrate the applicability of our neural networks.  
The effective action generated by the neural networks accurately approximates the target action with dynamical fermions.  
We show that the results of SLHMC are consistent with those of HMC.

Our network can be applied to general-purpose tasks in lattice QCD, even in cases with dynamical fermions.  
For example, it can be used to simulate overlap fermions via a generalized domain-wall action with adjustable parameters, enhancing overlap in reweighting techniques \cite{Fukaya:2013vka, Tomiya:2016jwr}.  
By mapping between two different probability distributions, our networks may enhance recently developed flow-based algorithms \cite{Albergo:2019eim, rezende2020normalizing, Kanwar:2020xzo, Boyda:2020hsi, Albergo:2021vyo}.  
Applications to classification tasks involving gauge-symmetric data are also of interest \cite{Favoni:2020reg, Matsumoto:2019jia}.

Note added.--Recently, we learned about the related works
Refs. \cite{Abbott:2022zhs,Abbott:2022zsh,Bacchio:2022vje,Lehner:2023bba,Aronsson:2023rli,Lehner:2023prf,Abbott:2023thq,Gerdes:2022eve,Gerdes:2024rjk}.
\section*{Acknowledgments}
The work of A.T. was supported by the RIKEN Special Postdoctoral Researcher program and partially by JSPS  KAKENHI Grant Numbers 20K14479, 22H05112, and 22H05111.
Y.N. was partially supported by JSPS- KAKENHI Grant Numbers 18K11345, 18K03552, 22K12052, 22K03539, 22H05111 and 22H05114. 
The calculations were partially performed using the supercomputing system HPE SGI8600 at the Japan Atomic Energy Agency.

Y.N. and A.T. contributed equally to this work.

\appendix 

\section{Basics of neural networks\label{sec:fcnn}}
Here we introduce conventional neural networks to be self-contained. Please refer \cite{Akinori2021DLAP} to see detail for example.
Let us take $\vec{x} \in \mathbb{R}^{n}$ is an input vector of a neural network.
A neural network is symbolically written as,
a composite function,
\begin{align}
f_\theta(\vec{x})
&=
\sigma_L(W_L
\sigma_{L-1}(W_{L-1}
\sigma_{L-2}(W_{L-2}
\cdots
\sigma_1(W_1 \vec{x})
\cdots
)
)
),
\label{eq:conventiona_neural_net}
\end{align}
where 
$W_l \in \mathbb{R}^{n_{l+1}
,n_l}$, $l=1,2,\cdots, L$
and $\sigma_k(\cdot)$ ($k=1,\cdots$) is an element-wise non-linear function.
We note $\theta$ as a set of parameters,
namely all elements of $W_l$ ($l=1,\cdots,L$), which 
determine to minimize a loss function. 
Note that, including convolutional neural networks, most of known neural networks can be written in this form.
It is known that, deep neural networks are universal approximator\cite{ zhou2018universality, kidger2020universal, tabuada2020universal, johnson2018deep}, namely which can approximate any maps in desired precision as Fourier expansion in a certain order.

Let us introduce notation for later purpose.
One of ingredients is a linear transformation, 
\begin{align}
z^{(l)}_i
= \sum_j w^{(l)}_{ij}
u^{(l-1)}_j,
\end{align}
and the other is a non-linear function,
${u}^{(l)}_i = \sigma_l(z^{(l)}_i)$.
We call $z_i^{(l)}$ as a pre-activation variable.
$w^{(l)}_{ij}$ is an element of $W_l$.
${u}^{(l)}_i$ is $i$-th component of $\vec{u}^{(l)}$ and $\vec{u}^{(0)}=\vec{x}$.
Both transformation is needed to realizes a universal  function from a vector space to another vector space \cite{ zhou2018universality, kidger2020universal, tabuada2020universal, johnson2018deep}.

To determine the parameters, we use a loss function.
A concrete form if the loss function is not
necessary but for example for a regression
we can take mean square error,
\begin{align}
L_\theta(\mathcal{D}) = \frac{1}{2|\mathcal{D}|} \sum_{(\vec{x}_i,\vec{y}_i)\in \mathcal{D}}
(\vec{y}_i-f_\theta(\vec{x}_i))^2,
\end{align}
where $\vec{y}_i$ is desired answer for $\vec{x}_i$,
desired output for $\vec{x}_i$.
$\mathcal{D}$ is a set of pairs of data,
which is replaced by a part of data for mini-batch training. 
$|\mathcal{D}|$ is the size of data.
This quantified a kind of distance between 
probability distribution of the answer and distribution of output. 
We can choose appropriate a loss function for each problem.
Popular choice is the Kullback-Leibler divergence (relative entropy), the cross-entropy or the square difference\cite{Akinori2021DLAP}.

A neural network is not simple convex function in general, so there are no best way to tune parameters to the optimal value.
Practically, parameters in neural networks are tuned (trained) by the  gradient descent,
\begin{align}
\theta \leftarrow \theta -\eta \frac{\partial L_\theta(\mathcal{D})}{\partial \theta},
\end{align}
where $\eta$ is a real small positive number called learning rate
and $\theta$ represents elements of weight matrices $W^{(l)}$.
This is called stochastic gradient descent since it uses sampling approximation to evaluate the value of gradient instead of the exact expectation value.
An extreme case, the size of $\mathcal{D}$ is taken to 1, it is called on-line training.

The derivative term
$
\frac{\partial L_\theta(\mathcal{D})}{\partial \theta}
$, 
can be evaluated by a recursive formula called backpropagation and the delta rule.
\begin{align}
\frac{\partial L_\theta(\mathcal{D})}{\partial w^{(l)}_{ij}}
=
\sum_k
\frac{\partial L_\theta(\mathcal{D})}{\partial z^{(l)}_k}
\frac{\partial z^{(l)}_k}{\partial w^{(l)}_{ij}}
=
\delta^{(l)}_i
u^{(l-1)}_j,
\end{align}
we defined,
\begin{align}
\delta^{(l)}_i = 
\frac{\partial L_\theta(\mathcal{D})}{\partial z^{(l)}_i}.
\end{align}
while $
z^{(l+1)}_i
= \sum_j w^{(l+1)}_{ij}
\sigma_l(z^{(l)}_j)
$, we can relate $\delta^{(l)}$
and $\delta^{(l+1)}$,
\begin{align}
\delta^{(l)}_i
&=
\sum_j
\frac{\partial L_\theta(\mathcal{D})}{\partial z^{(l+1)}_j}
\frac{\partial z^{(l+1)}_j}{\partial z^{(l)}_i}
=
\sum_k
\delta^{(l+1)}_k
\frac{\partial z^{(l+1)}_k}{\partial z^{(l)}_i} \nonumber \\
&=
\sum_k
\delta^{(l+1)}_k
 w^{(l+1)}_{ki}
\sigma'_l(z^{(l)}_i) .
\end{align}
These recursive equations enables us to calculate correction very efficiently.
The information propagates $\vec{x}=\vec{u}^{(1)}\to
\vec{u}^{(2)}\to\cdots \to \vec{u}^{(L)}$.
The correction propagate backwards,
\begin{align}
\vec{\delta}^{(1)}\leftarrow
\vec{\delta}^{(2)} \leftarrow
\cdots
\leftarrow
\vec{\delta}^{(L)}, 
\end{align}
where $\vec{\delta}^{(l)}$ has $\delta^{(l)}_k$ as elements.
This is called backpropagation or backprop in short because error propagates backwards.

\section{Connection to smearing}\label{sec:connect}
In this subsection, we identify relation between the covariant neural network and smearing schemes.

\subsection{Connection to APE and n-HYP smearing}
For APE-type smearing
\cite{Albanese:1987ds,Hasenfratz_2007},
the parameters and functions are taken as, 
\begin{align}
    w_1^{(l)} &= \alpha,\;\;w_2^{(l)} = \frac{\alpha}{6} \\
     {\cal G}_{\mu,n}( U^{(l)}) &= V_{\mu}^{\dagger (l)}[U](n) \label{eq:APE-g}
\end{align}
where 
${\cal N}(z)$ is a normalization function,
which guarantees output of the layer is an element of the gauge group (or  U(1) extended one with the normalization function).
To express n-HYP, ${\cal N}(z)$ has to be taken to a projection function
and coefficients be taken to appropriate ones.
$V_{\mu}^{\dagger (l)}[U](n) $ is a staple.

\subsection{Connection to stout smearing}
Smearing step in the stout smearing \cite{Morningstar_2004} is expressed as 
\begin{align}
    U_{\mu}^{(l+1)}(n) &= \exp ( Q_{\mu}^{(l)}(n)) U_{\mu}^{(l)}(n) \\
    &= U_{\mu}^{(l)}(n) + \left( \exp ( Q_{\mu}^{(l)}(n)) - 1 \right) U_{\mu}^{(l)}(n).
\end{align}
So, we can identify,
\begin{align}
    w_1^{(l)} &= w_2^{(l)} = 1 \label{eq:stout-w} \\
    {\cal N}(z) &= z \label{eq:stout-N}\\
     {\cal G}_{\mu,n}( U^{(l)}) &=  \left( \exp (Q_{\mu}^{(l)}(n)) - 1 \right) U_{\mu}^{(l)}(n)
     \label{eq:stout-g}
\end{align}
where $Q_{\mu}^{(l)}(n)$ is defined as 
\begin{align}
Q_{\mu}^{(l)}(n) = 2[\Omega_{\mu}^{(l)}(n)]_\text{TA}
\label{eq:stout-Q}
\end{align}
and $\Omega_{\mu}^{(l)}(n)$ is constracted by untraced plaquette. Here $\text{TA}$ means traceless-antihermitian operation.

Throughout this paper, we call the covariant neural network with 
Eq.~
\eqref{eq:APE-g}
as APE-type neural network,
and
with Eq.~
\eqref{eq:stout-w}
\eqref{eq:stout-N}
\eqref{eq:stout-g}
\eqref{eq:stout-Q}
as stout-type neural network.
As we will discuss later,
the stout-type neural network can contain trainable parameters in $Q_{\mu}^{(l)}(n)$.

\section{Parametrization for stout-type covariant neural network} \label{sec:stout}
For stout-type covariant neural network, one has to take $w_1^{(l)}= w_2^{(l)}=1$
otherwise $\mathcal{N}$ has to take as normalization function as n-HYP.
Weights can be included in $\Omega_\mu^{(l)}$.
For example we can choose as,
\begin{align}
\Omega^{(l)}_{\mu}(n) = 
\rho^{(l)}_\text{plaq} O^\text{plaq}_{\mu}(n) + 
\rho^{(l)}_\text{rect} O^\text{rect}_{\mu}(n) + 
\rho^{(l)}_\text{poly} O^\text{poly}_{\mu}(n),
\label{eq:omega_stout_train}
\end{align}
where 
$O^\text{plaq}_{\mu}(n)$,
$O^\text{rect}_{\mu}(n)$,  
$O^\text{poly}_{\mu}(n)$,
are
plaquette, rectangular loop, and Polyakov loop operators for $\mu$ direction originated from a point $n$, respectively. Every operators in $\Omega^{(l)}_{\mu}(n)$ is not traced as in the stout and HEX smearing \cite{Morningstar_2004, Capitani:2006ni}.

\section{Connection to fully connected neural network} \label{sec:full}
The covariant neural network fallback to
a conventional neural network
a linear function of a gauge field $A_\mu(n)$
if $a$ is small for $U_\mu(n)= \mathrm{e}^{\mathrm{i} a A_\mu(n)}
\approx 1+\mathrm{i} a A_\mu(n)$.
We follow the idea in \cite{Capitani:2006ni}, perturbative analysis of smearing.
For example, 
in the APE type smearing with a staple for plaquette is,
\begin{align}
&
w_1^{(l-1)}
U_{\mu}(n)
+
w_2^{(l-1)}
\sum_{\mu\neq\nu}
[
U_\nu(n+\hat\mu)
U^\dagger_\mu(n+\hat\nu)
U^\dagger_\nu(n) \nonumber \\
&+\text{opposite}],\\
&=
w_1^{(l-1)}
\mathrm{i} a A_{\mu}(n) 
+w_2^{(l-1)}
\sum_{\mu\neq\nu}
\big[
\mathrm{i} a A_\nu(n+\hat\mu)
-\mathrm{i} a A_\mu(n+\hat\nu)\notag\\&\;\;\;
-\mathrm{i} a A_\nu(n) 
+\text{(opposite)}\big]
+ \text{const}
+O(a^2) ,
\end{align}
where $\text{(opposite)}$ indicates a staple along with the opposite side.
This is nothing but a linear summation of input.
This is fed to an non-linear function,
which corresponds to an activation function.
Moreover, if we do not use weight sharing,
the right hand side of equation above is extended as,
\begin{align}
&w_{1,\mu}^{(l-1)}(n)
\mathrm{i} a A_{\mu}(n) 
+\sum_{\mu\neq\nu}
[
\mathrm{i} a 
w_{2,\nu}^{(l-1)} (n+\hat\mu)
A_\nu(n+\hat\mu) \notag \\
&\;\;\;\;
-\mathrm{i} a 
w_{2,\nu}^{(l-1)} (n+\hat\nu)
A_\mu(n+\hat\nu)  \notag \\
&\;\;\;\;
-\mathrm{i} a 
w_{2,\nu}^{(l-1)} (n)
A_\nu(n)
+\text{(opposite)}]
+ \text{const}
+O(a^2) .
\end{align}
If the gauge group is U(1), the gauge field $A_\mu(n)$ becomes real valued.
This can be regarded as a dense layer in a conventional neural network
if we identify a pair of index $(\mu,n)$
as index for a input vector in conventional fully connected layer.

\section{Rank-2 Delta rule and HMC force} \label{sec:rank2delta}
HMC is the {\it de facto} standard algorithm in lattice QCD because it can deal with dynamical fermions \cite{Duane:1987de}. HMC is based on the Metropolis algorithm and the molecular dynamics to obtain configurations. The molecular dynamics uses gradient of the action with respect to the gauge field, which is called HMC force. For smeared action, one has to treat smeared links as a composite function of un-smeared (= thin) links, which is calculated by the chain rule.
The HMC force with the covariant neural network is defined as 
\begin{align}
    F_{\mu,n}[U] 
    &= \left[U_\mu(n)\frac{\partial S[U^{(L)}]}{\partial U_{\mu}(n)} \right]_\text{TA}
    =
    \left[U_\mu(n) \delta_\mu^{(l=0)}(n) \right]_\text{TA},
\end{align}
where $U^{(L)}$ is the output from the covariant neural network and $L$ is the number of covariant layers.
The HMC force can be written in $ \delta_\mu^{(l=0)}(n)$, which is actually the same object with $\Sigma(n)$ in \cite{Morningstar_2004} for the stout smearing case.

With the star products and tensor version of chain rules, we can straightforwardly derive rank-2 version of the delta rule, which is a recurrence formula of $\delta_\mu^{(l)}(n)$. 
We note that there were similar derivations of HMC force with specific smearing schemes. 
To derive the recurrence formula of $\delta_\mu^{(l)}(n)$, we use the following chain rule,
\begin{align}
    \frac{\partial S}{\partial z_{\mu}^{(l)}(n)} &= \sum_{\mu',n'} \left[
      \frac{\partial S}{\partial z_{\mu'}^{(l+1)}(n')} \star   \frac{\partial z_{\mu'}^{(l+1)}(n')}{\partial z_{\mu}(n)} \right.
       \nonumber \\
      &+ \left. \frac{\partial S}{\partial z_{\mu'}^{(l+1)\dagger}(n')} \star   \frac{\partial z_{\mu'}^{(l+1)\dagger}(n')}{\partial z_{\mu}(n)} \right].
\end{align}
This is an extended Wiltinger derivative to complex matrix \cite{Kamleh:2004xk} and constraint of special unitary is taken care of.

On the final layer $l = L$, $\delta^{(L)}_{\mu}(n)$ becomes 
\begin{align}
    \delta^{(L)}_{\mu}(n) &= \sum_{\alpha=I,II}  \frac{\partial S_\theta}{\partial  U^{(L)\alpha}_{\mu}(n)} \star
     \frac{\partial {\cal N} [  z^{(L)}_{\mu}(n)]^{\alpha}}{\partial  z^{(L)}_{\mu}(n)}  ,
\end{align}
Here, we define $A^{I}(n) \equiv A$ and  $A^{II} \equiv A^{\dagger}$. 

On the $l$-th layer,  $\delta^{(l)\alpha}_{\mu}(n)$ is written as 
\begin{align}
  &   \delta^{(l)}_{\mu,\alpha}(n) 
    =  \sum_{\mu',m,\beta} 
     \delta^{(l+1)}_{\mu',\beta}(m)
     \star \frac{\partial z^{(l+1)\beta}_{\mu'}(m)}{\partial  z^{(l)\alpha}_{\mu}(n)}, \\
 &  
     = \sum_{\beta=I,II} \left\{ 
      w_1^{(l)} \delta^{(l+1)}_{\mu,\beta}(n)
     \star 
     \frac{\partial  
           {\cal N} [  z^{(l)\beta}_{\mu}(n)]
          }{\partial z^{(l)\alpha}_{\mu}(n) } \nonumber \right. \\
    & \left. 
    +  w_2^{(l)}   \sum_{\mu',m} 
      \delta^{(l+1)}_{\mu',\beta}(m) \star  \sum_{\gamma=I,II} 
       \frac{\partial  {\cal G}_{\mu',m}( U^{(l)} )^{\beta}}{\partial U^{(l)\gamma}_{\mu}(n) }
       \star  \frac{\partial  {\cal N} [  z^{(l)}_{\mu}(n)]^{\gamma}}{\partial z^{(l)\alpha}_{\mu}(n) }     
       \right\},
\end{align}
where $\alpha,\gamma = I, II$.
This is the rank-2 delta rule for the covariant neural network.
And by using this delta rule, we can optimize the weights in the network through a gradient optimizer.

We can derive a derivative with respect to $\rho$ using formula above, 
\begin{align}
\frac{\partial S}{\partial \rho^{(l)}_i}
     &= 2 \operatorname{Re}
     \sum_{\mu',m} 
     \operatorname{tr} \left[
 U^{(l)\dagger}_{\mu'}(m)  
        \Lambda_{\mu',m} 
        \frac{\partial C}{\partial \rho_i^{(l)}}
    \right],
\end{align}
where $\rho_i^{(l)}$ is a coefficient of a loop operator type $i$ (plaquette, rectangler, Polyakov loop, etc) in the level $l$.
$C$ is sum of staples for loop operators in $\Omega^{(l)}_\mu(n)$.
$\Lambda_{\mu,m}$ is defined in \cite{Morningstar_2004}.

\section{Tensor calculus and star product\label{sec:tensor}}
In this subsection we derive useful formulae based on \cite{Kamleh:2004xk} but we impose the star product match to the matrix chain rule, which enables us to derive exactly the same formula in \cite{Morningstar_2004} with the star product formulation.

\subsubsection*{Definition of matrix derivative}
The derivative of a scalar $f$ function of a matrix $A$ of independent variables, with respect to the matrix $A$ is a rank-2 tensor:
\begin{align}
    [\frac{\partial f[A]}{\partial A}]^i{}_j &\equiv 
    \frac{\partial f}{\partial A^j{}_i}.
\end{align}
The derivative of a matrix $M$ function of a matrix $A$ of independent variables, with respect to the matrix $A$ is a rank-4 tensor: 
\begin{align}
    [ \frac{\partial M[A]}{\partial A} ]^i{}_j{}^k{}_l &\equiv 
     \frac{\partial}{\partial A^j{}_k} M[A]^i{}_l
\end{align}
This is matrix-generalized Wirtinger derivative and for derivative with a complex number, it is fall back to conventional Wirtinger derivative.
The derivative of a matrix $M$ function of a scalar $x$ of independent variables, with respect to the matrix $x$ is a rank-2 tensor:
\begin{align}
    [\frac{\partial M[x]}{\partial x}]^i{}_j &\equiv 
    \frac{\partial M[x]^i{}_j}{\partial x}.
\end{align}

\subsubsection*{Chain rules and star products}
The chain rule is given as 
\begin{align}
    [\frac{\partial f[A]}{\partial A}]^i{}_j &= 
    \frac{\partial f}{\partial A^j{}_i} 
    = \sum_{kl}  \frac{\partial f}{\partial B^k{}_l}  \frac{\partial B^k{}_l}{\partial A^j{}_i} \\
    &= \sum_{kl}  [ \frac{\partial f}{\partial B}]^l{}_k [ \frac{\partial B}{\partial A}]^k{}_j{}^i{}_l
\end{align}
Then, by defining rank-2-rank4 ``star''-product
\begin{align}
    [A \star T]^i{}_j &\equiv \sum_{kl}  A^l{}_k T^k{}_j{}^i{}_l,
\end{align}
we have 
\begin{align}
      \frac{\partial f}{\partial A}  &= 
      \frac{\partial f}{\partial B} \star  \frac{\partial B}{\partial A}
\end{align}
We consider the following chain rule:
\begin{align}
       [ \frac{\partial M[A]}{\partial A} ]^i{}_j{}^k{}_l &=   \sum_{nm} \frac{\partial  M[A]^i{}_l }{\partial B^n{}_m}  \frac{\partial  B^n{}_m }{\partial A^j{}_k} \\
       &=  \sum_{nm} [ \frac{\partial M[B]}{\partial B}]^i{}_n{}^m{}_l [ \frac{\partial B[A]}{\partial A}]^n{}_j{}^k{}_m
\end{align}
By defining rank-4-rank4 star product:
\begin{align}
    (S \star T)^i{}_j{}^k{}_l &=  \sum_{nm} S^i{}_n{}^m{}_l T^n{}_j{}^k{}_m
\end{align}
we have 
\begin{align}
      \frac{\partial M[A]}{\partial A} &=
       \frac{\partial M[B]}{\partial B} \star  \frac{\partial B[A]}{\partial A}
\end{align}
We consider a different chain rule:
\begin{align}
    \frac{\partial f}{\partial x} &= 
    \sum_{ij}  \frac{\partial f}{\partial A^j{}_i}  \frac{\partial A^j_i}{\partial x} \\
    &= \sum_{ij} \left[   \frac{\partial f}{\partial A }\right]^i{}_j \left[\frac{\partial A}{\partial x} \right]^j{}_i \\
    &= {\rm Tr} \left[
     \left[   \frac{\partial f}{\partial A }\right]
      \left[\frac{\partial A}{\partial x} \right]
     \right]
     \end{align}
     Here, $f$ and $x$ are scalars. 
     We can define  a rank-2-rank-2 star product. 
     \begin{align}
 A \star B   &\equiv {\rm Tr} \left[
 A B
 \right]
\end{align}
And we have 
\begin{align}
    \frac{\partial f}{\partial x} &= 
    \sum_{ij}  \frac{\partial f}{\partial A^j{}_i}  \frac{\partial A^j_i}{\partial x} \\
    &= \sum_{ij} \sum_{kl}
     \frac{\partial f}{\partial B^k{}_l} 
      \frac{\partial  B^k{}_l}{\partial  A^j_i} \frac{\partial A^j_i}{\partial x} \\
      &=\sum_{ij} \sum_{kl}
[ 
 \frac{\partial f}{\partial B}
]{}^l{}_k [
 \frac{\partial B}{\partial A}
]{}^k{}_i{}^j{}_l[
 \frac{\partial A}{\partial x}
]{}^i{}_j \\
&= 
\sum_{ij} \sum_{kl}
[ 
 \frac{\partial f}{\partial B}
]{}^l{}_k [
 \frac{\partial B}{\partial A}
 \star 
 \frac{\partial A}{\partial x}
]{}^k{}_l \\
&=  \frac{\partial f}{\partial B} \star (
 \frac{\partial B}{\partial A}
 \star 
 \frac{\partial A}{\partial x}
)
     \end{align}
Here we use the following definition: 
\begin{align}
    [S \star A]^i{}_j &\equiv S^i{}_k{}^l{}_jA^k{}_l
\end{align}
This is a definition of a rank-4-rank-2 star product.

\subsection{Product rules}
We consider the following product rule: 
\begin{align}
     [\frac{\partial (f[A] M[A])}{\partial A}]^i{}_j{}^k{}_l &=  \frac{\partial}{\partial A^j{}_k} (f[A] M[A]^i{}_l]) \\
     &=  \frac{\partial f[A]}{\partial A^j{}_k}  M[A]^i{}_l + f[A]  \frac{\partial}{\partial A^j{}_k} M[A]^i{}_l \\
     &= [\frac{\partial f[A]}{\partial A}]^k{}_j  M[A]^i{}_l +   f[A] [ \frac{\partial M[A]}{\partial A} ]^i{}_j{}^k{}_l
\end{align}
By defining ``direct'' product:
\begin{align}
(A \oplus B )^i{}_j{}^k{}_l &= A^k{}_j B^i{}_l
\end{align}
we have 
\begin{align}
     \frac{\partial f[A] M[A]}{\partial A} 
     &= \frac{\partial f[A]}{\partial A} \oplus M[A] +   f[A] \frac{\partial M[A]}{\partial A}
\end{align}

We consider the matrix $M = ABC$. The derivative is 
\begin{align}
     [ \frac{\partial M[B]}{\partial B} ]^i{}_j{}^k{}_l &=  \frac{\partial}{\partial B^j{}_k} M[B]^i{}_l \\
     &= \sum_{n,m} \frac{\partial}{\partial B^j{}_k} (A^i{}_n B^n_{m} C^{m}_l) \\
     &  \sum_{n,m} (A^i{}_n\delta_{jn} \delta_{km} C^{m}_l) \\
     &=   A^i{}_j C^{k}_l 
\end{align}
By defining the outer product
\begin{align}
    (A \otimes B)^i{}_j{}^k{}_l &\equiv  A^i{}_j B^{k}_l 
\end{align}
we have 
\begin{align}
      \frac{\partial M[B]}{\partial B} &=  A \otimes C
\end{align}

We consider 
\begin{align}
  &  [ \frac{\partial (Y[A]Z[A])}{\partial A} ]^i{}_j{}^k{}_l =
     \frac{\partial}{\partial A^j{}_k} [YZ]^i{}_l \\
     &=   \sum_n \frac{\partial}{\partial A^j{}_k} (Y^i{}_n Z^n{}_l) \\
     &= \sum_n \frac{\partial Y^i{}_n }{\partial A^j{}_k}  Z^n{}_l +  \sum_n Y^i{}_n \frac{\partial  Z^n{}_l }{\partial A^j{}_k} \\
     &= \sum_n [ \frac{\partial Y[A]}{\partial A} ]^i{}_j{}^k{}_n  Z^n{}_l + \sum_n  Y^i{}_n [ \frac{\partial Z[A]}{\partial A} ]^n{}_j{}^k{}_l 
\end{align}
By defining the contraction: 
\begin{align}
    [AT]^i{}_j{}^k{}_l &= \sum_n A^i{}_n T^n{}_j{}^k{}_l  \\
    [TA]^i{}_j{}^k{}_l &= \sum_n T^i{}_j{}^k{}_n A^n{}_l
\end{align}
we have 
\begin{align}
     \frac{\partial YZ}{\partial A} &=  \frac{\partial Y[A]}{\partial A} Z + Y \frac{\partial Z[A]}{\partial A} 
\end{align}

\subsection{Useful formulas}
The useful formulas are given as 
\begin{align}
\frac{\partial A}{\partial A} &= I \otimes I \\
           A \star (B \oplus C) 
     &= {\rm Tr} (AC) B \\
       A \star (B \otimes C)&= 
      C A B \\
       (A \otimes B) \star (C \otimes D) &= (AC) \otimes (DB) \\
       (A \oplus B) \star (C \oplus D) &=  {\rm Tr}(AD) (C \oplus B) \\
        (A \otimes B) \star (C \oplus D) &= C \oplus (ADB) \\
        (A \oplus B) \star (C \otimes D) &= (DAC) \oplus B \\
         S \star (T \star K)  &=   (S \star T) \star K \\
         A \star (T \star K)  &= (A \star T) \star K\\
         (A \otimes B)C &= A \otimes (BC) \\
         (A \oplus B)C &= A \oplus (BC) \\
          C(A \otimes B) &= (CA) \otimes B\\
           C(A \oplus B) &= A \oplus (CB)\\
           (S \star T)C &= (SC) \star T \\
           A \star (TC) &= (CA) \star T \\
           S \star (I \otimes I) &= S \\
             B^{\dagger} \star \frac{\partial M^{\dagger}}{\partial A^{\dagger}} &= [B \star \frac{\partial M}{\partial A}]^{\dagger} 
\end{align}

\section{Backpropagation for stout smearing}\label{sec:stout}
\subsection{Definition}

Smearing step in the stout smearing \cite{Morningstar_2004} is expressed as 
\begin{align}
    U_{\mu}^{(l+1)}(n) &= \exp ( Q_{\mu}^{(l)}(n)) U_{\mu}^{(l)}(n) \\
    &= U_{\mu}^{(l)}(n) + \left( \exp ( Q_{\mu}^{(l)}(n)) - 1 \right) U_{\mu}^{(l)}(n).
\end{align}
So, we can identify,
\begin{align}
    w_1^{(l)} &= w_2^{(l)} = 1 \\
    {\cal N}(z) &= z \\
     {\cal G}_{\mu,n}( U^{(l)}) &=  \left( \exp (Q_{\mu}^{(l)}(n)) - 1 \right) U_{\mu}^{(l)}(n)
\end{align}
where $Q_{\mu}^{l}(n)$ and 
\begin{align}
Q_{\mu}(n) &= -\frac{1}{2}(\Omega_{\mu}(n) - \Omega_{\mu}^{\dagger}(n)) + \frac{1}{2N} \operatorname{Tr} (\Omega_{\mu}^{\dagger}(n) -\Omega_{\mu}(n) ) \\
\Omega_{\mu}(n) &= C_{\mu}(n)U_{\mu}^{\dagger}(n)
\end{align}
and $C_{\mu}(n)$ is the weighted sum of the perpendicular staples which begin at lattice site $n$ and terminate at neighboring site $n+\hat{\mu}$.
For example, the staple of a plaquette loop is given as 
\begin{align}
    &\sum_{\nu \neq \mu} \rho_{\mu,\nu} \left(
    U_{\nu}(n)U_{\mu}(n+\hat{\nu})U_{\nu}^{\dagger}(n+\hat{\mu}) \nonumber \right. \\
    &+ \left. U_{\nu}^{\dagger}(n-\hat{\nu})U_{\mu}(x-\nu)U_{\nu}(x-\hat{\nu} + \hat{\mu})  
    \right)
\end{align}
\subsection{Backpropagation}
In the stout smearing scheme, the $\delta_{\mu}(n)$ is given as 
\begin{align}
    \delta_{\mu}^{(l)}(n) &= \frac{\partial S}{\partial z_{\mu}^{(l)}(n)} = 
    \frac{\partial S}{\partial U_{\mu}^{(l)}(n)} \\
     \bar{\delta}_{\mu}^{(l)}(n) &=  \frac{\partial S}{\partial U_{\mu}^{(l)\dagger}(n)} =
      \frac{\partial S}{\partial U_{\mu}^{(l)}(n)} \star 
       \frac{\partial  U_{\mu}^{(l)}(n)}{\partial U_{\mu}^{(l)\dagger}(n)} \\
       &= 
        \frac{\partial S}{\partial U_{\mu}^{(l)}(n)} \star 
       \frac{\partial  (U_{\mu}^{(l)\dagger}(n))^{-1}}{\partial U_{\mu}^{(l)\dagger}(n)}\\
       &= -
        \frac{\partial S}{\partial U_{\mu}^{(l)}(n)} \star 
        (U_{\mu}^{(l)}(n) \otimes U_{\mu}^{(l)(n))})\\
        &= -U_{\mu}^{(l)}(n)   \delta_{\mu}^{(l)}(n)
         U_{\mu}^{(l)}(n)  
\end{align}
On the $l$-th layer, the backpropagation is given as 
 \begin{align}
&     \delta^{(l)}_{\mu}(n) 
=  \delta^{(l+1)}_{\mu}(n)
     \star 
     (I \otimes I)
  \nonumber \\
& + \sum_{\mu',m} 
  \delta^{(l+1)}_{\mu'}(m) \star 
 \left[ 
     \frac{\partial  {\cal G}_{\mu',m}( U^{(l)} )}{\partial U^{(l)}_{\mu}(n) } 
     \star 
     (I \otimes I)
\right] \nonumber \\
&-   \sum_{\mu',m} 
(U_{\mu',m}   \delta^{(l+1)}_{\mu'}(m) U_{\mu',m})
 \star 
 \left[ 
     \frac{\partial  {\cal G}_{\mu',m}( U^{(l)} )^{\dagger}}{\partial U^{(l)}_{\mu}(n) } 
     \star 
    (I \otimes I)
      \right]
\end{align}
By substituting 
\begin{align}
 \frac{\partial  {\cal G}_{\mu',m}( U^{(l)} )}{\partial U^{(l)}_{\mu}(n) } &= 
   \frac{\partial \exp (Q_{\mu'}^{(l)}(m))}{\partial U^{(l)}_{\mu}(n) }  U_{\mu'}^{(l)}(m) \nonumber \\
& +  (\exp (Q_{\mu'}^{(l)}(m))-I)  \frac{ \partial U_{\mu'}^{(l)}(m) }{\partial U^{(l)}_{\mu}(n) } \\
 &= 
   \frac{\partial \exp (Q_{\mu'}^{(l)}(m))}{\partial U^{(l)}_{\mu}(n) }  U_{\mu'}^{(l)}(m) \nonumber \\
& +  (\exp (Q_{\mu'}^{(l)}(m))-I)  \delta_{\mu,\mu'} \delta_{n,m} I \otimes I  \\
  \frac{\partial  {\cal G}_{\mu',m}^{\dagger}( U^{(l)} )}{\partial U^{(l)}_{\mu}(n) } &= 
  U_{\mu'}^{(l)\dagger}(m)  \frac{\partial \exp (Q_{\mu'}^{(l)\dagger}(m))}{\partial U^{(l)}_{\mu}(n) } 
\end{align}
we have 
 \begin{align}
&     \delta^{(l)}_{\mu}(n) 
= 
 \delta^{(l+1)}_{\mu}(n))  \exp (Q_{\mu'}^{(l)}(m))
     \star 
     (I \otimes I) 
 \nonumber \\
&+ \sum_{\mu',m} 
 ( U_{\mu'}^{(l)}(m)  \delta^{(l+1)}_{\mu'}(m)) \star 
  \frac{\partial \exp (Q_{\mu'}^{(l)}(m))}{\partial U^{(l)}_{\mu}(n) } 
     \star 
     (I \otimes I) \nonumber \\
     &- \sum_{\mu',m} 
 (  \delta^{(l+1)\dagger}_{\mu'}(m) U_{\mu'}^{(l)\dagger}(m) ) \star 
 \left[ 
 - \frac{\partial \exp (Q_{\mu'}^{(l)\dagger}(m))}{\partial U^{(l)}_{\mu}(n) }  \right]
     \star 
     (I \otimes I)
\end{align}
Here we use $\delta = -U^{\dagger} \delta^{\dagger} U^{\dagger} $.

The exponentials are given as 
\begin{align}
   \frac{\partial \exp (Q_{\mu'}^{(l)}(m))}{\partial U^{(l)}_{\mu}(n) } &= 
    \frac{\partial \exp ( Q_{\mu'}^{(l)}(m))}{\partial Q^{(l)}_{\mu'}(m) } \star  \frac{
    \partial Q^{(l)}_{\mu'}(m)
    }{\partial U^{(l)}_{\mu}(n) } \\
    &=  \frac{\partial \exp ( Q_{\mu'}^{(l)}(m))}{\partial Q^{(l)}_{\mu'}(m) } \star 
      \left[
       \frac{
    \partial Q^{(l)}_{\mu'}(m)
    }{\partial \Omega^{(l)}_{\mu}(n) } \star 
     \frac{
    \partial \Omega^{(l)}_{\mu'}(m)
    }{\partial U^{(l)}_{\mu}(n) } \right. \nonumber \\
   & \left. +       \frac{
    \partial Q^{(l)}_{\mu'}(m)
    }{\partial \Omega^{(l)\dagger}_{\mu}(n) } \star
     \frac{
    \partial \Omega^{(l)\dagger}_{\mu'}(m)
    }{\partial U^{(l)}_{\mu}(n) }
    \right]      \\
      &=  \frac{\partial \exp ( Q_{\mu'}^{(l)}(m))}{\partial Q^{(l)}_{\mu'}(m) } \star 
        \frac{
    \partial Q^{(l)}_{\mu'}(m)
    }{\partial \Omega^{(l)}_{\mu}(n) }
    \star
      \left[
     \frac{
    \partial \Omega^{(l)}_{\mu'}(m)
    }{\partial U^{(l)}_{\mu}(n) } \right. \nonumber \\
&  \left.  -
     \frac{
    \partial \Omega^{(l)\dagger}_{\mu'}(m)
    }{\partial U^{(l)}_{\mu}(n) }
    \right] 
\end{align}
and 
\begin{align}
   \frac{\partial \exp (Q_{\mu'}^{(l)\dagger}(m))}{\partial U^{(l)}_{\mu}(n) } 
      &=  \frac{\partial \exp ( Q_{\mu'}^{(l)\dagger}(m))}{\partial Q^{(l)\dagger}_{\mu'}(m) } \star 
        \frac{
    \partial Q^{(l)\dagger}_{\mu'}(m)
    }{\partial \Omega^{(l)}_{\mu}(n) }
    \star
      \left[
     \frac{
    \partial \Omega^{(l)}_{\mu'}(m)
    }{\partial U^{(l)}_{\mu}(n) } \right. \nonumber \\
& \left.  -
     \frac{
    \partial \Omega^{(l)\dagger}_{\mu'}(m)
    }{\partial U^{(l)}_{\mu}(n) }
    \right] 
\end{align}
By using the fact that $Q^{(l)}$ is a traceless anti-hermitian matrix, we have 
\begin{align}
      \frac{
    \partial Q^{(l)\dagger}_{\mu'}(m)
    }{\partial \Omega^{(l)}_{\mu}(n) } &= 
    -  \frac{
    \partial Q^{(l)}_{\mu'}(m)
    }{\partial \Omega^{(l)}_{\mu}(n) }
\end{align}
Thus, the equations can be expressed as 
\begin{align}
           &    \delta^{(l)}_{\mu}(n) 
= 
 \delta^{(l+1)}_{\mu}(n) \exp (Q_{\mu'}^{(l)}(m))
     \star 
     (I \otimes I) 
 \nonumber \\
&+ \sum_{\mu',m} 
 ( U_{\mu'}^{(l)}(m)  \delta^{(l+1)}_{\mu'}(m)) \star 
  \frac{\partial \exp ( Q_{\mu'}^{(l)}(m))}{\partial Q^{(l)}_{\mu'}(m) } \star 
        \frac{
    \partial Q^{(l)}_{\mu'}(m)
    }{\partial \Omega^{(l)}_{\mu}(n) }  \nonumber \\
&     \star
      \left[
     \frac{
    \partial \Omega^{(l)}_{\mu'}(m)
    }{\partial U^{(l)}_{\mu}(n) }  -
     \frac{
    \partial \Omega^{(l)\dagger}_{\mu'}(m)
    }{\partial U^{(l)}_{\mu}(n) }
    \right] 
     \star 
     (I \otimes I) \nonumber \\
     &- \sum_{\mu',m} (U_{\mu'}^{(l)}(m)  \delta^{(l+1)}_{\mu'}(m) )^{\dagger}
 \star 
 \frac{\partial \exp ( Q_{\mu'}^{(l)\dagger}(m))}{\partial Q^{(l)\dagger}_{\mu'}(m) } \star 
        \frac{
    \partial Q^{(l)}_{\mu'}(m)
    }{\partial \Omega^{(l)}_{\mu}(n) } \nonumber \\
   &  \star
      \left[
     \frac{
    \partial \Omega^{(l)}_{\mu'}(m)
    }{\partial U^{(l)}_{\mu}(n) }
  -
     \frac{
    \partial \Omega^{(l)\dagger}_{\mu'}(m)
    }{\partial U^{(l)}_{\mu}(n) }
    \right] 
     \star 
     (I \otimes I) \\
     &= 
 \delta^{(l+1)}_{\mu}(n) \exp (Q_{\mu'}^{(l)}(m))
     \star 
     (I \otimes I) 
 \nonumber \\
&+ \sum_{\mu',m} 
(M_{\mu',m} - M_{\mu',m}^{\dagger})
 \star 
        \frac{
    \partial Q^{(l)}_{\mu'}(m)
    }{\partial \Omega^{(l)}_{\mu}(n) } \nonumber \\
    &    \star
      \left[
     \frac{
    \partial \Omega^{(l)}_{\mu'}(m)
    }{\partial U^{(l)}_{\mu}(n) }
  -
     \frac{
    \partial \Omega^{(l)\dagger}_{\mu'}(m)
    }{\partial U^{(l)}_{\mu}(n) }
    \right] 
     \star 
     (I \otimes I),\\
&   M_{\mu',m} \equiv  ( U_{\mu'}^{(l)}(m)  \delta^{(l+1)}_{\mu'}(m)) \star 
  \frac{\partial \exp ( Q_{\mu'}^{(l)}(m))}{\partial Q^{(l)}_{\mu'}(m) }   
\end{align}
By using the following equation: 
\begin{align}
     \frac{\partial  Q_{\mu}(y)}{\partial \Omega_{\mu}(y)} &= -\frac{\partial }{\partial \Omega_{\mu}(y)}   \left[ \frac{1}{2} \left[\Omega_{\mu}(y)^{\dagger} -  \Omega_{\mu}(y) \right] \right. \nonumber \\
     & \left. - \frac{1}{2N} {\rm Tr} \Omega_{\mu}(y)^{\dagger} I +  \frac{1}{2N} {\rm Tr}  \Omega_{\mu}(y) I \right] \\
     &=
     \frac{1}{2} \frac{\partial  \Omega_{\mu}(y)}{\partial \Omega_{\mu}(y)}
     -  \frac{1}{2N} 
     \frac{\partial  
     {\rm Tr}  \Omega_{\mu}(y)
     }{\partial \Omega_{\mu}(y)}  
      \oplus I \\
      &= \frac{1}{2} I \otimes I
     -  \frac{1}{2N} I  \oplus I 
\end{align}
and the following formula
\begin{align}
     A \star ( \frac{1}{2} I \otimes I
     -  \frac{1}{2N} I  \oplus I) &= 
        \frac{1}{2} A \star (I \otimes I)
     -  \frac{1}{2N} A \star (I  \oplus I) \\
     &=  \frac{1}{2} A - \frac{1}{2N} {\rm Tr} (A) I
\end{align}
we obtain 
\begin{align}
      & \delta^{(l)}_{\mu}(n) = 
 \delta^{(l+1)}_{\mu}(n) \exp (Q_{\mu'}^{(l)}(m))
     \star 
     (I \otimes I) 
 \nonumber \\
&+ \sum_{\mu',m} 
  \Lambda_{\mu',m} 
    \star
      \left[
     \frac{
    \partial \Omega^{(l)}_{\mu'}(m)
    }{\partial U^{(l)}_{\mu}(n) }
  -
     \frac{
    \partial \Omega^{(l)\dagger}_{\mu'}(m)
    }{\partial U^{(l)}_{\mu}(n) }
    \right] 
     \star 
     (I \otimes I) \\
     & = 
 \delta^{(l+1)}_{\mu}(n) \exp (Q_{\mu'}^{(l)}(m)) \nonumber \\
&+ \sum_{\mu',m} 
  \Lambda_{\mu',m} 
    \star
      \left[
     \frac{
    \partial \Omega^{(l)}_{\mu'}(m)
    }{\partial U^{(l)}_{\mu}(n) }
  -
     \frac{
    \partial \Omega^{(l)\dagger}_{\mu'}(m)
    }{\partial U^{(l)}_{\mu}(n) }
    \right] 
\end{align}
Here, $\Lambda_{\mu,n}$ is defined as 
\begin{align}
    \Lambda_{\mu',m} &\equiv 
      \frac{1}{2} (M_{\mu',m} - M_{\mu',m}^{\dagger}) - \frac{1}{2N} {\rm Tr} (M_{\mu',m} - M_{\mu',m}^{\dagger}) I
\end{align}
With the use of the following equations:
\begin{align}
      \frac{
    \partial \Omega^{(l)}_{\mu'}(m)
    }{\partial U^{(l)}_{\mu}(n) } &=   \frac{
    \partial( C_{\mu',m} U_{\mu'}^{\dagger}(m))
    }{\partial U^{(l)}_{\mu}(n) } =  \frac{
    \partial C_{\mu',m} 
    }{\partial U^{(l)}_{\mu}(n) }  U^{(l)\dagger}_{\mu'}(m) \\
\frac{
    \partial \Omega^{(l)\dagger}_{\mu'}(m)
    }{\partial U^{(l)}_{\mu}(n) } &=   \frac{
    \partial(  U_{\mu'}(m) C_{\mu',m}^{\dagger})
    }{\partial U^{(l)}_{\mu}(n) } \nonumber \\
&    =  U_{\mu'}(m)  \frac{
    \partial C_{\mu',m}^{\dagger} 
    }{\partial U^{(l)}_{\mu}(n) } +  \frac{
    \partial U_{\mu'}(m)
    }{\partial U^{(l)}_{\mu}(n) }  C_{\mu',m}^{\dagger} \\
    &=  U_{\mu'}(m)  \frac{
    \partial C_{\mu',m}^{\dagger} 
    }{\partial U^{(l)}_{\mu}(n) } + \delta_{\mu,\mu'} \delta_{n,m} I \otimes I C_{\mu,n}^{\dagger} \\
    &= U_{\mu'}(m)  \frac{
    \partial C_{\mu',m}^{\dagger} 
    }{\partial U^{(l)}_{\mu}(n) } + \delta_{\mu,\mu'} \delta_{n,m} I \otimes C_{\mu,n}^{\dagger} \\
     \frac{
    \partial \Omega^{(l)}_{\mu'}(m)
    }{\partial U^{(l)}_{\mu}(n) } -
    \frac{
    \partial \Omega^{(l)\dagger}_{\mu'}(m)
    }{\partial U^{(l)}_{\mu}(n) } &=
     \frac{
    \partial C_{\mu',m} 
    }{\partial U^{(l)}_{\mu}(n) }  U^{(l)\dagger}_{\mu}(n) \nonumber \\
&    - U_{\mu'}(m)  \frac{
    \partial C_{\mu',m}^{\dagger} 
    }{\partial U^{(l)}_{\mu}(n) } - \delta_{\mu,\mu'} \delta_{n,m} I \otimes C_{\mu,n}^{\dagger}
\end{align}
We finally obtain 
\begin{align}
      & \delta^{(l)}_{\mu}(n) = 
 \delta^{(l+1)}_{\mu}(n) \exp (Q_{\mu'}^{(l)}(m))
 - \Lambda_{\mu,n} \star ( I \otimes C_{\mu,n}^{\dagger})
 \nonumber \\
 &
+ \sum_{\mu',m} 
  \Lambda_{\mu',m} 
    \star
      \left[
        \frac{
    \partial C_{\mu',m} 
    }{\partial U^{(l)}_{\mu}(n) }  U^{(l)\dagger}_{\mu'}(m) 
    - U_{\mu'}(m)  \frac{
    \partial C_{\mu',m}^{\dagger} 
    }{\partial U^{(l)}_{\mu}(n) }
    \right]  \\
    &= 
     \delta^{(l+1)}_{\mu}(n) \exp (Q_{\mu'}^{(l)}(m))
 -C_{\mu,n}^{\dagger} \Lambda_{\mu,n}
 \nonumber \\
 &
+ \sum_{\mu',m} 
  \Lambda_{\mu',m} 
    \star
      \left[
        \frac{
    \partial C_{\mu',m} 
    }{\partial U^{(l)}_{\mu}(n) }  U^{(l)\dagger}_{\mu'}(m) 
    - U_{\mu'}(m)  \frac{
    \partial C_{\mu',m}^{\dagger} 
    }{\partial U^{(l)}_{\mu}(n) }
    \right] \\
 &= 
     \delta^{(l+1)}_{\mu}(n) \exp (Q_{\mu'}^{(l)}(m))
 -C_{\mu,n}^{\dagger} \Lambda_{\mu,n}
 \nonumber \\
 &
+ \sum_{\mu',m} 
      \left[(U^{(l)\dagger}_{\mu'}(m)  
        \Lambda_{\mu',m} 
      ) \star 
        \frac{
    \partial C_{\mu',m} 
    }{\partial U^{(l)}_{\mu}(n) }  
    -
    ( 
        \Lambda_{\mu',m}  U_{\mu'}(m)
      ) \star   \frac{
    \partial C_{\mu',m}^{\dagger} 
    }{\partial U^{(l)}_{\mu}(n) }
    \right]   \\
     &= 
     \delta^{(l+1)}_{\mu}(n) \exp (Q_{\mu'}^{(l)}(m))
 -C_{\mu,n}^{\dagger} \Lambda_{\mu,n}
 \nonumber \\
 &
+ \sum_{\mu',m} 
      \left[(U^{(l)\dagger}_{\mu'}(m)  
        \Lambda_{\mu',m} 
      ) \star 
        \frac{
    \partial C_{\mu',m} 
    }{\partial U^{(l)}_{\mu}(n) }  
    +
    (  U_{\mu'}^{\dagger}(m)
        \Lambda_{\mu',m} 
      )^{\dagger} \star   \frac{
    \partial C_{\mu',m}^{\dagger} 
    }{\partial U^{(l)}_{\mu}(n) }
    \right]  
\end{align}
The above equation has only one star product. 
The rank-4 tensors 
 $ \frac{
    \partial C_{\mu',m} 
    }{\partial U^{(l)}_{\mu}(n) }  $ and $  \frac{
    \partial C_{\mu',m}^{\dagger} 
    }{\partial U^{(l)}_{\mu}(n) }$ are usually expressed as 
\begin{align}
     \frac{
    \partial C_{\mu',m} 
    }{\partial U^{(l)}_{\mu}(n) } &\equiv \sum_i A_{\mu,m}^i \otimes B_{\mu,m}^i \\
       \frac{
    \partial C_{\mu',m}^{\dagger} 
    }{\partial U^{(l)}_{\mu}(n) } &\equiv \sum_i \bar{A}_{\mu,m}^i \otimes \bar{B}_{\mu,m}^i
\end{align}
Then, we have 
\begin{align}
      & \delta^{(l)}_{\mu}(n) 
     = 
     \delta^{(l+1)}_{\mu}(n) \exp (Q_{\mu'}^{(l)}(m))
 -C_{\mu,n}^{\dagger} \Lambda_{\mu,n}
 \nonumber \\
 &
+ \sum_{\mu',m} \sum_i \left[
 B_{\mu,m}^i U^{(l)\dagger}_{\mu'}(m)  
        \Lambda_{\mu',m}  A_{\mu,m}^i \right. \nonumber \\
&\left.         -
   \bar{B}_{\mu,m}^i 
    \Lambda_{\mu',m}  U_{\mu'}(m)
      \bar{A}_{\mu,m}^i 
    \right]  
\end{align}
This equation is equivalent to the original one derived without tensor calculations. 

\subsection{parameter derivative}
The derivative with respect to the parameter $\rho$ can be calculated as 
\begin{align}
    \frac{\partial S}{\partial \rho^{(l)}} &= 
    \sum_{\mu,n} \left[
      \frac{\partial S}{\partial z_{\mu}^{(l)}(n)} 
      \star 
          \frac{\partial z_{\mu}^{(l)}(n)}{\partial \rho^{(l)}}
          +  \frac{\partial S}{\partial z_{\mu}^{(l)\dagger}(n)} 
      \star 
          \frac{\partial z_{\mu}^{(l)\dagger}(n)}{\partial \rho^{(l)}}
          \right]\\
          &= 
       \sum_{\mu,n} \left[
       \delta_{\mu}(n) 
       \star (
       \frac{\partial \exp Q_{\mu}^{(l)}(n)}{\partial \rho^{(l)}} U_{\mu}(n) ) \right. \nonumber \\
& \left.       +    \bar{\delta}_{\mu}(n) 
       \star 
       (U_{\mu}^{\dagger}(n)
       \frac{\partial \exp Q_{\mu}^{(l)\dagger}(n)}{\partial \rho^{(l)}})
       \right] \\
       &= 
        \sum_{\mu,n} 
      \left[(U^{(l)\dagger}_{\mu}(n)  
        \Lambda_{\mu,n} 
      ) \star 
        \frac{
    \partial C_{\mu,n} 
    }{\partial \rho^{(l)} }  \right. \nonumber \\
&\left.     +
    (  U_{\mu}^{\dagger}(n)
        \Lambda_{\mu,n} 
      )^{\dagger} \star   \frac{
    \partial C_{\mu,n}^{\dagger} 
    }{\partial \rho^{(l)}}
    \right] \\
    &= 
       2 \sum_{\mu,n} \operatorname{Re} \operatorname{Tr}
      \left[U^{(l)\dagger}_{\mu}(n)  
        \Lambda_{\mu,n} 
        \frac{
    \partial C_{\mu,n} 
    }{\partial \rho^{(l)} }  
    \right]
\end{align}
Here, we use $\frac{
    \partial C_{\mu,n}^{\dagger} 
    }{\partial \rho^{(l)}} = [\frac{
    \partial C_{\mu,n} 
    }{\partial \rho^{(l)}}]^{\dagger}$.
    



\subsection{Calculation of $\partial \exp(Q)/\partial Q$}
According to the Ref.~\cite{Morningstar_2004}, we have 
\begin{align}
\frac{\partial \exp(iQ)}{\partial \tau}&={\rm Tr}(Q \frac{\partial Q}{\partial \tau})B_1 + {\rm Tr}(Q^2 \frac{\partial Q}{\partial \tau})B_2 \nonumber \\
& + f_1 \frac{\partial Q}{\partial \tau} + f_2 \frac{\partial Q}{\partial \tau} Q + f_2 Q \frac{\partial Q}{\partial \tau},
\end{align}
in SU(3) system. 
Here, the coefficients are given in Ref.~\cite{Morningstar_2004}. 
The derivatives are related as 
\begin{align}
    \left[\frac{\partial A}{\partial Q} \right]{}^i{}_j{}^k{}_l  = \frac{\partial A^i{}_l}{\partial  Q^j{}_k} = \frac{\partial  A^i{}_l}{\partial  \tau}\frac{\partial \tau}{\partial  Q^j{}_k}.
\end{align}
By using following equation: 
\begin{align}
   &  \left[ \frac{\partial \exp(iQ)}{\partial \tau} \right]^i{}_l=Q^a{}_b \frac{\partial Q^b{}_a}{\partial \tau} B_1{}^i{}_l + [Q^2]^a{}_b \frac{\partial Q^b{}_a}{\partial \tau}B_2{}^i{}_l \nonumber \\
    &+ f_1 \frac{\partial Q{}^i{}_l }{\partial \tau} + f_2 \frac{\partial Q{}^i{}_a}{ \partial \tau} Q^a{}_l + f_2 Q^i{}_a \frac{\partial Q^a{}_l}{\partial \tau},
\end{align}
we have
\begin{align}
    & \left[ \frac{ \partial \exp(iQ)}{\partial \tau} \right]^i{}_l \frac{\partial \tau}{\partial Q^j{}_k} = Q^k{}_jB_1{}^i{}_l + [Q^2]^k{}_j B_2{}^i{}_l  + f_1 \delta_{ij} \delta_{lk} \nonumber \\
    &+ f_2 \delta_{ij} Q^k{}_l + f_2 Q^i{}_j  \delta_{lk}.
\end{align}
Then, we obtain 
\begin{align}
    \frac{\partial \exp(iQ)}{\partial Q} &= Q \oplus B_1 + Q^2 \oplus B_2 \nonumber \\
    &+ f_1 I \otimes I + f_2 I \otimes Q + f_2 Q \otimes I .
\end{align}
We also have the following equation: 
\begin{align}
    C \star \frac{\partial \exp(iQ)}{\partial  Q} &= {\rm Tr}(C B_1) Q  + {\rm Tr}(C B_2) Q^2  \nonumber \\
    & + f_1 C  + f_2 Q C + f_2 C Q.
\end{align}

In SU(N) system, $\partial \exp(Q)/\partial Q$ is expressed as 
\begin{align}
\frac{\partial \exp Q}{\partial Q} = \frac{\partial }{\partial Q} \sum_{n=0}^{\infty} \frac{1}{n!} Q^n = \sum_{n=0}^{\infty}  \frac{1}{n!}\sum_{k=0}^{n-1} Q^{k} \otimes Q^{n-1-k}.
\end{align}
Then, we have 
\begin{align}
    A \star \frac{\partial \exp Q}{\partial Q} = \sum_{n=0}^{\infty}  \frac{1}{n!}\sum_{k=0}^{n-1} Q^{n-1-k} A Q^k. 
\end{align}

\section{Mass dependence of the trainable parameters}\label{sec:mass}
\begin{figure*}[th]
\begin{center}
\includegraphics[scale=0.5]{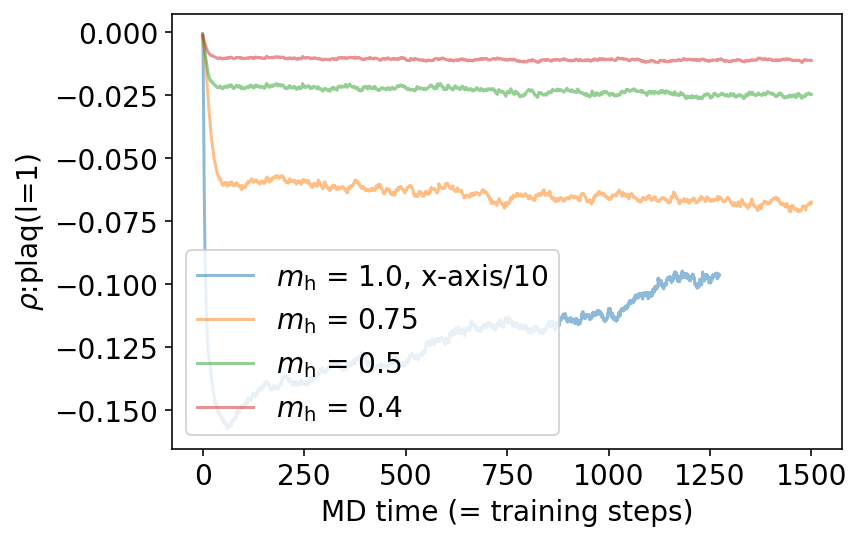} 
\includegraphics[scale=0.5]{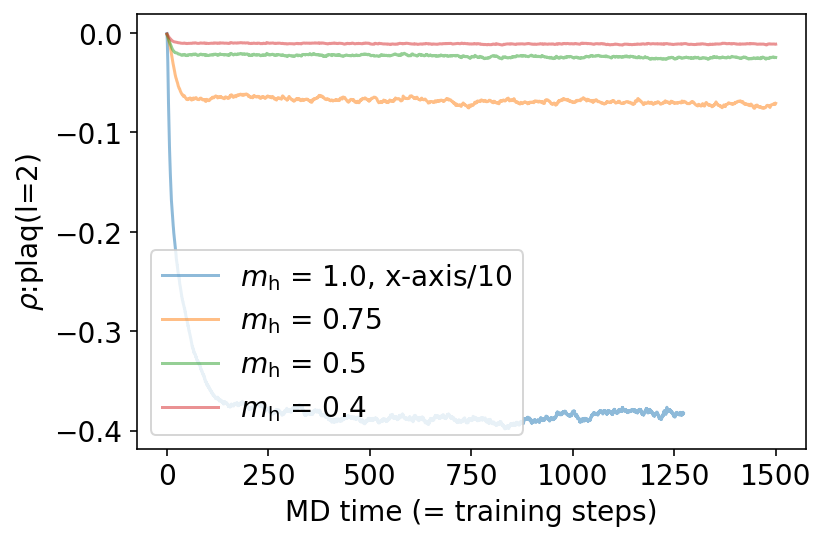}\\ 
\includegraphics[scale=0.5]{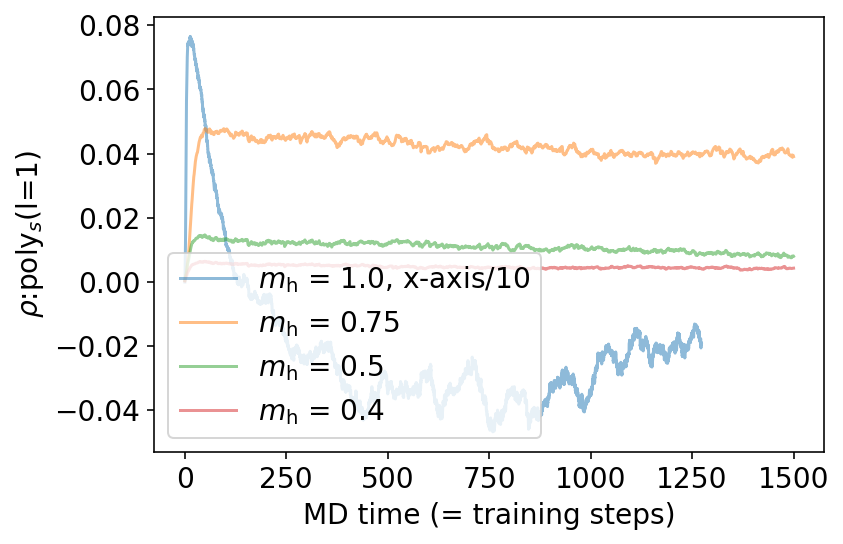} 
\includegraphics[scale=0.5]{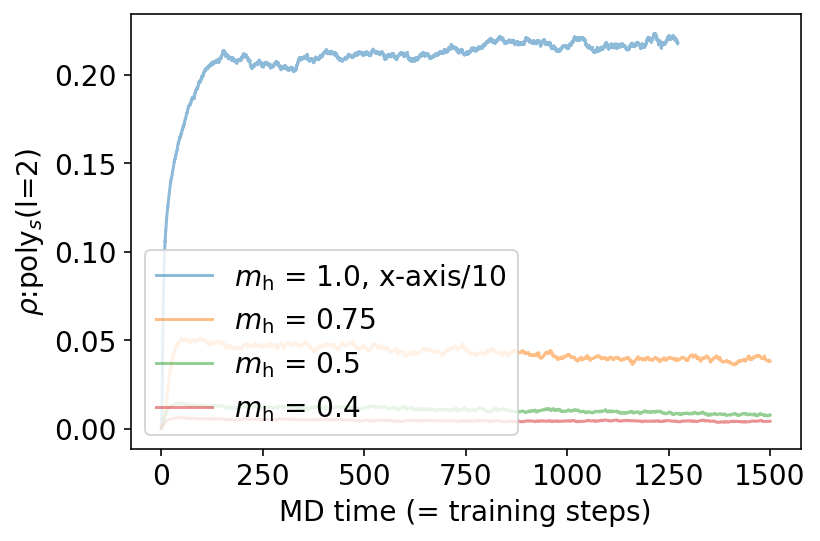}\\ 
\includegraphics[scale=0.5]{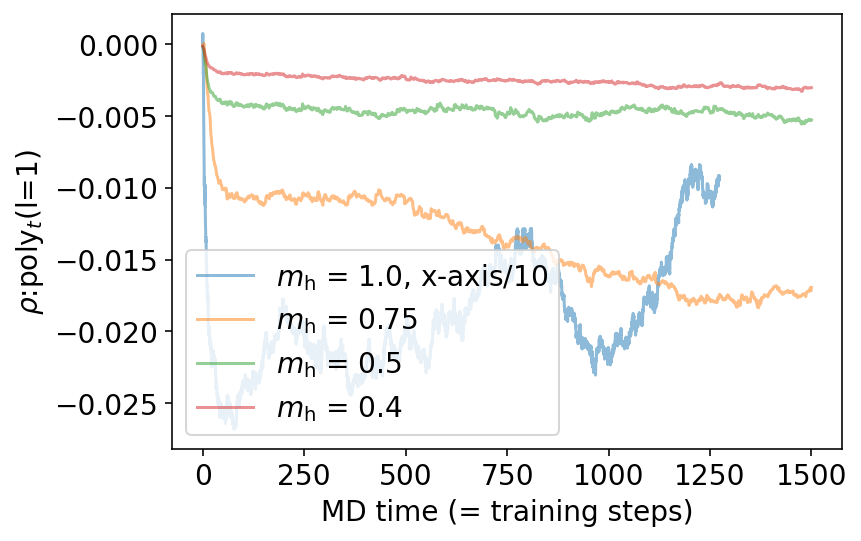} 
\includegraphics[scale=0.5]{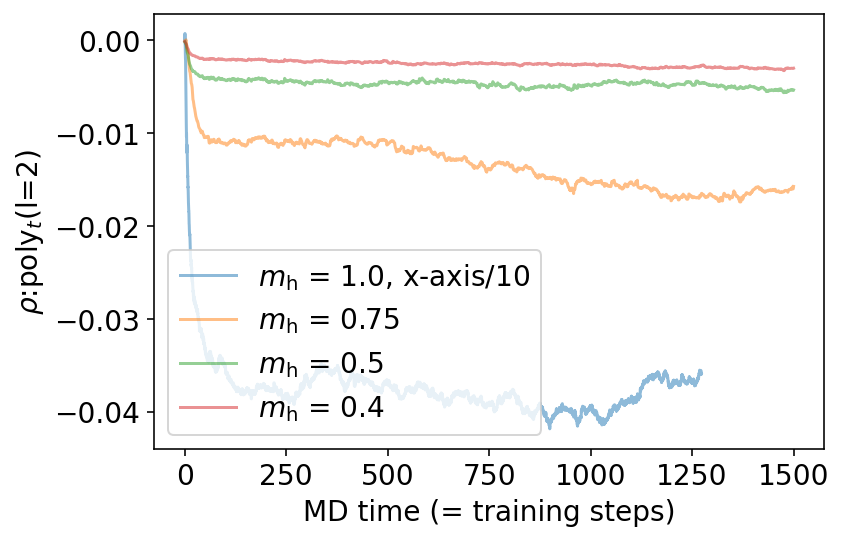} 
\end{center}
\caption{
Training history of weights in the neural network.
From top to bottom,
coefficients of plaquette, Polyakov loop for spatial directions, and Polyakov loop for temporal directions are displayed.
Left panels and right panels are weights in the first layer and the second layer.
\label{fig:rho_history}}
\end{figure*}

Fig. \ref{fig:rho_history} shows 
training history of weights in the neural network in HMC. From top to bottom,
coefficients of plaquette, Polyakov loop for spatial directions, and Polyakov loop for the temporal direction are displayed.
Left panels and right panels are weights for the first layer and the second layer.
One can see that all of weights $\rho\neq0$ for after the training.
All of weights in the first layer show same tendency with ones in the second layer.
This means, some of weights are irrelevant and we can reduce the degrees of freedom by adding a regulator term into the loss function.
Coefficients of Polyakov loop for spatial direction has opposite tendency with  temporal ones.
the weight for Polyakov loop with $m_\text{h}=1.0$ at $l=1$ has short plateau MD time $=1200$ and it goes down again around $1800$.
This indicates that, if we employ an effective action far from the target action, longer training steps are needed.
In SLHMC run, we use trained weight in $1500$-th step (Tab. \ref{tab:trained_weight_m04}).
Some of values are negative, while coefficients in conventional smearing are positive.

\begin{table}[htb]
\centering
\begin{tabular}{cc|c}
Layer& Loop & Value of $\rho$  \\\hline\hline
1&Plaquette & $-0.011146476388409423$\\
2&Plaquette & $-0.011164492428633698$\\
1&Spatial Polyakov loop & $-0.0030283193221172216$\\
2&Spatial Polyakov loop &$ -0.0029984533773388094$\\
1&Temporal Polyakov loop &  $0.004248021727233112$\\
2&Temporal Polyakov loop & $0.004195253380373369$
\end{tabular}
\caption{Trained weights for the current setup.
These values are used in the SLHMC run.
\label{tab:trained_weight_m04}
}
\end{table}

\bibliography{ref}


\end{document}